\documentclass[12pt,a4paper]{revtex4}
\usepackage{multirow}
\usepackage{amssymb}
\usepackage{amsmath,graphicx}
\usepackage{array}
\usepackage{braket}
\usepackage{bm}
\usepackage{enumerate}
\usepackage{epstopdf}
\usepackage{amsmath}
\usepackage{dcolumn,caption,booktabs}
\usepackage{slashed}
\usepackage{amsfonts}
\usepackage{adjustbox}
\usepackage{tcolorbox}
\usepackage{float}
\usepackage{collectbox}
\usepackage{listings}
\usepackage[unicode]{hyperref}
\usepackage[mathscr]{eucal}
\usepackage{graphics}
\usepackage{subcaption}
\newcommand{\hs}{\hspace*{0.3cm}}

\newcommand{\be}{\begin{equation}}
\newcommand{\ee}{\end{equation}}
\newcommand{\bea}{\begin{eqnarray}}
\newcommand{\eea}{\end{eqnarray}}
\newcommand{\ben}{\begin{enumerate}}
	\newcommand{\een}{\end{enumerate}}
\newcommand{\bde}{\begin{widetext}}
	\newcommand{\ede}{\end{widetext}}
\newcommand{\nn}{\nonumber}
\newcommand{\crn}{\nonumber \\}

\newcommand{\al}{\alpha}
\newcommand{\la}{\lambda}

\newcommand{\fr}{\frac}

\newcommand{\bc}{\begin{center}}
	\newcommand{\ec}{\end{center}}
\newcommand{\Ga}{\Gamma}

\newcommand{\ep}{\epsilon}

\newcommand{\La}{\Lambda}

\newcommand{\Om}{\Omega}
\begin{document}	
	\title{\boldmath Baryogenesis and gravitational waves in the Zee-Babu model}
\author{Vo Quoc Phong$^{a,b}$}
	\email{vqphong@hcmus.edu.vn}
	\affiliation{$^a$ Department of Theoretical Physics, University of Science, Ho Chi Minh City 700000, Vietnam\\
	$^b$ Vietnam National University, Ho Chi Minh City 700000, Vietnam}	
	\author{Nguyen Chi Thao$^{a,b}$}
	\email{thaynguyenchithao@gmail.com}
	\affiliation{$^a$ Department of Theoretical Physics, University of Science, Ho Chi Minh City 700000, Vietnam\\
$^b$	Vietnam National University, Ho Chi Minh City 700000, Vietnam}
	\author{Hoang Ngoc Long$^{c,d}$}
	\email{hoangngoclong@tdmu.edu.vn (corresponding author)}
\affiliation{$^c$ Center for Forecasting Studies,  Thu Dau Mot University, Binh Duong Province, Vietnam
\\
$^c$ Institute of Physics, Vietnam Academy of Science and Technology, 10 Dao Tan, Ba Dinh, 10000 Hanoi, Vietnam}
	\date{\today }

\begin{abstract}
To explain the matter-antimatter asymmetry in the Zee-Babu (ZB) model, the sphaleron process in the baryogenesis scenario is calculated. It always satisfies the de-coupling condition and the strength of phase transition ($S$) is always greater than $1$ in the presence of triggers other than that in the  Standard Model, which are singly ($h^{\pm}$) and doubly  ($k^{\pm\pm}$) charged scalar bosons. Sphaleron energies are in the range of 5-10 TeV, in calculation with bubble profiles containing free parameters and assuming nuclear bubbles of $h^{\pm}$ and $k^{\pm\pm}$ are very small. We tested the scaling law of sphaleron again with an average error of $10\%$. When the temperature is close to the critical one ($T_c$), the density of nuclear bubble is produced very large and decreases as the temperature decreases. The key parameter is $\al $ which results in the gravitational wave density parameter ($\Om h^2$) in the range of $10^{-14}$ to $10^{-12}$ when $\beta/H^*=22.5$, this is not enough to detect gravitational waves from electroweak phase transition (EWPT) according to the present LISA data but may be detected in the future. As the larger strength of phase transition is, the more $\al $ increases (this increase is almost linear with $S$), the larger the gravitational wave density parameter is. Also in the context of considering the generation of gravitational waves, in the ZB  model we calculated $\al \sim \text{a few} \times 10^{-2}\ll 1$, so rigorously conclude that the EWPT is not strong even though $S>1$. We also suggest that, for a model with a lot of extra scalar particles and particles which play a role in mass generation, the stronger the EWPT process and the larger $\Om h^2$ can be.
\end{abstract}
\pacs{11.15.Ex, 12.60.Fr, 98.80.Cq}
\maketitle
Keywords:  Spontaneous breaking of gauge symmetries,
	Extensions of electroweak Higgs sector, Particle-theory models (Early Universe)	

\tableofcontents
\newpage
\section{Introduction}

The Standard Model (SM) has set up numerous great comes about that concur with tests and gave us a clear system of how matter interatomic with each other. The foundation of the SM is the concept of symmetries such as Lorentz and gauge symmetries.  The gauge symmetry unavoidably leads to the presence of gauge bosons which mediate interactions. Initially, gauge bosons are massless which are not suitable for weak interaction acting in short range.
Within the Higgs component, which is fundamentally changes for the situation, can produce masses spontaneously, simultaneously keeps the SM renormalizability.

However, there are a few crucial issues that have not been unraveled by the SM and one of them is the matter-antimatter asymmetry problem and we need the hypothesis to be steady with gravity. We need a dynamical clarification for this and have to examine electroweak stage prepare that happens within the early universe.  To explain this problem, besides many others,  two standard scenarios  are leptogenesis and baryogenesis. In the case of leptogenesis, according to Ref.~\cite{72}, the typical scale is much higher than electroweak scale. In 1967, Sakharov proposed three conditions that a model must have in arrange to illuminate the baryogenesis issue \cite{sakharov}, which are baryon number violation, C and CP violation (via sphaleron process), and out-of-equilibrium condition, ie., the first order EWPT and the last condition should be considered first. In any case, EWPT inside the setting of the SM does not offer great arrangements since it does not fulfill adequately the two final   conditions of baryogenesis. Hence, models beyond SM got to be taken into consideration. A common choice is to treat the SM as a complete symmetric foundation and add only extraneous particles. Of course, we need to come up with good reasons for adding new particles.

In addition, since the Higgs boson was found at the LHC, Physics Particle nearly completed its mission that give a more precise understanding of mass. The EWPT issue will be one of the pressing issues, has been calculated in the SM as in Refs.\cite{mkn,SME,SMEb,SMEc,SMEd,michela} and in beyond SM or in numerous other settings as in Refs.\cite{2,2b,2c,BSM,BSMb,BSMc,majorana,majoranab,thdm,thdmb,ESMCO,elptdm,elptdma,elptdmb,elptdmc,elptdmd,phonglongvanb,phonglongvan2,SMS,dssm,munusm,lr,singlet,singletb,singletc,singletd,mssm1,mssm1b,mssm1c,twostep,twostepb,twostepc,1101.4665,1101.4665b,1101.4665c,jjgb,jjgc,jjgd,Ahriche1,Ahriche2,Ahriche2b,Ahriche3,Ahriche3b,Ahriche3c,Ahriche3d,Ahriche3e,Fuyuto,Fuyutob,Fuyutoc,span,chr,cde,kusenko}. We summarize most generally that the triggers for the EWPT process are either particles outside the SM or manually added quantities \cite{2,2b,2c,BSM,BSMb,BSMc,majorana,majoranab,thdm,thdmb,ESMCO,elptdm,elptdma,elptdmb,elptdmc,elptdmd,phonglongvanb,phonglongvan2,SMS,dssm,munusm,lr,singlet,singletb,singletc,singletd,mssm1,mssm1b,mssm1c,twostep,twostepb,twostepc,1101.4665,1101.4665b,1101.4665c,jjgb,jjgc,jjgd,Ahriche2,Ahriche2b,Ahriche3,Ahriche3b,Ahriche3c,Ahriche3d,Ahriche3e,Fuyutob,Fuyutoc,span,chr,cde,kusenko}. A commonplace negative solution is that the EWPT's strength satisfies the decoupling condition at a few hundred GeV scale with the mass of Higgs boson must be less than $125$ GeV \cite{mkn,SME,SMEb,SMEc,SMEd,michela}. So far, the nature of a first-order EWPT can be non-SM-like particles \cite{2,2b,2c,majorana,majoranab,thdm,thdmb,ESMCO,elptdm,elptdma,elptdmb,elptdmc,elptdmd,phonglongvanb,phonglongvan2,epjc,zb,singlet,singletb,singletc,singletd,mssm1,mssm1b,mssm1c,twostep,twostepb,twostepc,chiang3}. Another curiously finding is that EWPT is independent of the gauge \cite{zb, 1101.4665, 1101.4665b,1101.4665c,Arefe}. The damping impact within the self-energy term or daisy loops have little contributions. When considering daisy loops, the change of effective potential leads to a diminishment of the strength of EWPT generally by 2/3 \cite{r23}. Subsequently, the contributions do not make an enormous alter to the strength of EWPT or, in other words, it is not the origin of EWPT.

By ensuring existence C and CP violations, in the context of topological transitions, the third condition of baryogenesis is defined as $\Ga_{sph}\sim \mathcal{O}(T) \textrm{exp}\left(\fr{- E_{sph}}{T}\right) \ll H_{rad}$ \cite{spha-huble,decoupling, decouplingb, decouplingc, decouplingd}, where $\Ga_{sph}$ and $E_{sph}$ are the sphaleron rate and energy, respectively, $H_{rad}$ is the Hubble expansion rate in the radiation-dominated period, and $\mathcal{O}(T)\sim T^4$. This is known as the decoupling condition of sphaleron. In the SM, this condition is often expressed as $v_c/T_c>1$ by using an approximation  $E_{sph}(T)\approx [v(T)/v] E_{sph}(T=0)$ \cite{47, Ahriche1, Ahriche2,Ahriche2b}; however, in beyond SM models this approximation should be again considered with caution.

Sphaleron has also been calculated in various frameworks. We see there are two main directions: the first is to investigate the phenomena of sphaleron in models \cite{phu1spha,phu2spha,phu3spha,phu4spha,10,11,13,hagi,47,Ahriche1, Ahriche2,Ahriche2b,gauge0}, the second is to investigate by simulation and numerical methods \cite{michela,cosmotran1b}. We found that there are two basic ways to calculate the sphaleron energy \cite{plk}. The first way, from the sphaleron energy functional with spherically symmetric ansatzs, deriving the equations of motion (EOMs), and
solving these EOMs and substituting them into the sphaleron energy functional, we will get desired results. The second way, assuming the profile functions in the spherically symmetric ansatzs, and substituting them into the energy functionals,  then we will  get desired results. In the first way, we have to use the numerical method twice and have to check the convergence of the energy functional. In the second way, we only use the numerical method once, but assuming the profile functions must be very careful so that the energy functionals converge.

Another aspect, in General Relativity, stars, black holes \dots are highly concentrated masses of matter and there are supermassive black hole merger events, or collisions between two black holes. These collisions are so strong that they create distortions in space-time, resulting in their energy densities also changing, which changes the state of space-time and travels out into space at the speed of light. The propagation of curvature is called a gravitational wave \cite{1gw}. In addition, gravitational waves can be generated by core collapse of supernovae, the first order EWPT, oscillations of cosmic strings, inflation. In particular, the strong first order EWPT is one of the most promising of all these sources \cite{2gw,2gwa,2gwb}.

There have been many studies on gravitational waves in models \cite{phu1gw,phu2gw,phu2gwb,phu3gw,phu4gw,phu5gw,phu6gwsimulation,phu7gw,phu8gw,2gw,2gwa,2gwb,3gw,4gw,7gw, 7gwb, 8gw,9gw, 9gwb} as well as simulations \cite{phu6gwsimulation}. Most models give  the values of gravitational wave density parameter smaller than the observable value (about $10^{-10}$ \cite{7gw, 7gwb}). However, we see that the value of this parameter may be consistent with the observation if the latent heat parameter is about $1$.

Among the extended SM models, there are models closest to the SM which are The Two Double Higgs model, SMEFT, the left-right model,... But the ZB model is pretty simple because the symmetrical structure and particle sectors are like the SM.

In the ZB model, two extra charged scalars $h^{\pm}$ and $k^{\pm\pm}$
are added to the Higgs content. The kind of new scalars can play an important role in the early Universe. As shown in Refs.\cite{zeebabu, zeebabub, bzee, bzeea, bzeeb}, they can also be a reason for tiny mass of neutrinos through two loop or three-loop corrections.

The EWPT  in the ZB  model was investigated by us in Ref.\cite{zb}. In the mentioned paper, the strength  of electroweak phase transition ($S$) is greater than one. We initially concluded that the EWPT in the model under consideration is strong. However, this is not complete.

To complete the baryogenesis scenario and fully evaluate the EWPT process with the addition of gravitational waves, are the main purposes of this paper, which is organized as follows. In section \ref{sec2}, we summarize the ZB model and the EWPT with the one-loop effective potential. In section \ref{sec3}, electroweak sphalerons are calculated at zero and high temperature and we derive again the sphaleron scaling law. In section \ref{sec4}, $\al $ and $\Om h^2$ is calculated with different strength of phase transition. Finally, we summarize and make outlooks in section \ref{sec5}.

\section{Review on the Zee-Babu model and the first-order EWPT}\label{sec2}

In the ZB model, by adding two charged scalar bosons $h^{\pm}$ and $k^{\pm\pm}$ \cite{zeebabu, zeebabub}, the Lagrangian becomes
\bea\label{lzb}
\mathcal{L}  &=& L_{SM}+f_{ab}\overline{\psi_{aL}^c}\psi_{bL}h^{+} +h_{ab}^{'}\overline{l_{aR}^c}l_{bR}k^{++} + V(\phi,h,k)
\crn
& &+ (D_{\mu}h^+)^{\dagger}(D^{\mu}h^{+}) +
(D_{\mu}k^{++})^{\dagger}(D^{\mu}k^{++})+H.c\,,
\label{1}
\eea
where $\psi_L$ stands for the left-handed lepton doublet and $l_R$ for the right-handed lepton singlet. $h^{+}$ ia a singly charged boson and $k^{++}$ is a doubly charged boson. $a,b$ are the generation indices. The most important component is the Higgs sector which has four couplings between $h^{\pm}$ or $k^{\pm\pm}$ and SM-like Higgs boson \cite{zeebabu}:
\bea
V(\phi,h,k)&=&\mu^2\phi^{\dagger}\phi + u^2_1{\vert{h}\vert}^2 + u_2^2{\vert{k}\vert}^2 +\la  (\phi^{\dagger}\phi)^2 +
\la _h{ \vert{h}\vert}^4 + \la _k{ \vert{k}\vert}^4
\crn
& &  +\la _{hk}{ \vert{h}\vert}^2{ \vert{k}\vert}^2 + 2p^2{\vert{h}\vert}^{2} \phi^{\dagger}\phi
+ 2q^2{\vert{k}\vert}^{2} \phi^{\dagger}\phi + (\mu_{hk} h^{2} k^{++} + H.c)\, ,
\label{2}
\eea
where
\be
\phi=\left(\begin{array}{c}
	\rho^+ \\
	\rho^0
\end{array}\right)
\label{3}
\ee
and $v_0$ is a Vacuum Expectation Value (VEV) of $\rho^0$,
\be
\rho^{0}=\fr{1}{\sqrt{2}}\left( v_0+\sigma+i\zeta\right)\, .\label{4}
\ee

The masses of $h^\pm$ and $k^{{\pm\pm}}$ have two components, one comes from the interaction with the SM-like Higgs boson, the others are  free components ($u_1^2|h|^2$ or $u_2^2|k|^2$),
\bea\label{mass}
M^2_{h^\pm} &= &p^2v^2_0+u^2_1=m^2_{h^\pm}+ u^2_1,m^2_{h^\pm}=p^2v^2_0\, ,\crn
M^2_{{k^{\pm\pm}}}&=&q^2v^2_0+u^2_2=m^2_{{k^{\pm\pm}}}+u^2_2,m^2_{{k^{\pm\pm}}}=q^2v^2_0\, .\label{masshk}
\eea

Diagonalizing matrices in the kinetic component of the Higgs potential and neglect Goldstone bosons, we obtain
\begin{gather}\label{eq:SMfieldDepmasses}
\begin{aligned}
m_H^2(v_0)&=2\la  v^2_0\,,
\end{aligned}
\begin{aligned}
m_Z^2(v_0)&=\textstyle\fr{1}{4}(g^2+g'^2) v^2_0=A^2v^2_0\,,
\end{aligned}
\begin{aligned}
m_W^2(v_0)&=\textstyle\fr{1}{4}g^2 v^2_0=B^2v^2_0.
\end{aligned}
\end{gather}

The components $u_1$ and $u_2$ in Eq.~(\ref{mass}) do not participate in the EWPT or the effective potential because they are not components which are derived from the interaction between Higgs 
 and $h^{\pm}$ or $k^{\pm\pm}$ bosons. This can be seen clearly in Ref.~\cite{phonglongvan2}. Therefore, notice that in the next sections, the masses of $h^\pm$ and $k^{\pm\pm}$ 
 appearing hereafter
  are just $m_{h^\pm}$ and $ m_{k^{\pm\pm}}$.

There are usually two ways to calculate the effective potential, which is well summarized in Ref.~\cite{plk}. We recall the form of the one-loop effective potential without daisy loops in the Landau gauge \cite{zb}:
\bea\label{pf}
V _{eff}(v)&= &V _{0}(v)+\fr{3}{64\pi^2}\left( m _{Z}^{4}(v)ln\fr{m _{Z}^{2}(v)}{Q^2}
+2m _{W}^{4}(v)ln\fr{m _{W}^{2}(v)}{Q^2} - 4m _{t}^{4}(v)ln\fr{m _{t}^{2}(v)}{Q^2}\right)\crn
&+& \fr{1}{64\pi^2}\left(2m_{h^\pm}^{4}(v)ln\fr{m_{h^{\pm}}^{2}(v)}{Q^2} + 2m_{k^{\pm\pm}}^{4}(v){ln}\fr{m_{k^{\pm\pm}}^{2}(v)}{Q^2}+ m _ {H}^{4}(v)ln\fr{m_{H}^{2}(v)}{Q^2} \right)\crn
&+& \fr{3 T^4}{4\pi^2} \left\{F_{-} (\fr{m_Z(v)}{T})+F_{-}(\fr{m_W(v)}{T}) + 4F_{+}(\fr{m_t(v)}{T})\right\}\crn
&+& \fr{T^4}{4\pi^2}\left\{2{F}_{-}(\fr{m_{h^{\pm}}(v)}{T})+2{F}_{-}(\fr{m_{k^{\pm\pm}}(v)}{T}) + {F}_{-}(\fr{{m}_{H}(v)}{T})\right\}\, ,
\eea
where $v$ is the variable that changes with temperature, and at $0K$ then $v \equiv v_0=246$ GeV. $F_{\pm}$ are generalized integrals:
\bea
F_{\pm}\left(\fr{m_\phi}{T}\right) & = &\int_{0}^{\fr{m_\phi}{T}}\al  J_{\mp}^{1} (\al ,0)d\al,\crn
J_{\mp}^{1} (\al ,0)&=& 2\int_{\al } ^{\infty}\fr{(x^2-\al ^ 2)^{\fr{1}{2}}}{e^x\mp 1}dx.\label{kt8}
\eea

After full high-temperature expansion,  with an approximation of $F_{\pm}$, we get:
\be
V_{eff}(v)=D(T^2-T^2_0)v^2-ET|v|^3+\fr{\la _T}{4}v^4\,.
\label{EP-v-11}
\ee
The coefficients in the above potential have the following form:
\bea
D &=&\fr{1}{24 {v_0}^2} \left[
6 m_W^2(v_0) +3m_{Z_1}^2(v_0)
+m_{H}^2(v_0)+2m_{h^\pm}^2(v_0) +2m_{k^{\pm\pm}}^2(v_0) +6 m_t^2 (v_0)
\right],\crn
T_0^2&=&\fr{1}{D}\left\{\fr{m_{H}^2(v_0)}{4}
-\fr{1}{32\pi^2v_0^2}\left(6m_W^4(v_0)+3m_{Z_1}^4(v_0)+m_{H}^4(v_0)
\right.\right.\crn
&&\left.\left.\qquad \qquad \qquad \qquad+2m_{h^\pm}^4(v_0) +2m_{k^{\pm\pm}}^4(v_0) -12 m_t^4 (v_0)
\right)\right\},\crn
E &=& \fr{1}{12 \pi v_0^3} \left(
6 m_W^3(v_0) +3m_{Z_1}^3(v_0)
+m_{H}^3(v_0) +2m_{h^\pm}^3(v_0) +2m_{k^{\pm\pm}}^3(v_0)
\right),\label{D-T-E-lamda}\\
\la _T &=&
\fr{m_{H}^2(v_0)}{2 v_0^2}\left\{
1- \fr{1}{8\pi^2 v_0^2 (m_{H}^2(v_0))}\left[
6 m_W^4(v_0) \ln \fr{m_W^2(v_0)}{a_b T^2} 		
\right.\right.\crn
&&\qquad \left.\left.
+3m_{Z_1}^4(v_0) \ln\fr{m_{Z_1}^2(v_0)}{a_b T^2}+m_{H^0}^4(v_0) \ln \fr{m_{H^0}^2(v_0)}{a_b T^2}		
\right.\right.\crn
&&\qquad \left.\left.
+2m_{k^{\pm\pm}}^4(v_0)\ln\fr{m_{k^{\pm\pm}}^2(v_0)}{a_bT^2}+2m_{h^\pm}^4(v_0)\ln\fr{m_{h^\pm}^2(v_0)}{a_bT^2}-12 m_t^4(v_0)\ln\fr{m_t^2(v_0)}{a_F T^2}
\right] \right\}.\nn \eea

Nevertheless, the $5\%$ difference is between the full temperature-dependent integrals in Eq.~(\ref{kt8}) and their high-temperature expansion at order $T^2$ in $m/T$ when $\fr{m}{T}<2.2$ \cite{5percent}. Therefore,  according to our full estimation, $S$ cannot increase to $2.4$, is also in the range of $2.28$.

The potential in Eq.(\ref{EP-v-11}) has two minima:
\be \label{v-0}
v=0, \quad v \equiv v_c=\fr{2ET_c}{\la _{T_c}},
\ee
where $v_c$ is VEV at the critical temperature ($T_c$), which is defined:
\be \label{Tc}
T_c=\fr{T_0}{\sqrt{1-E^2/D\la _{T_c}}} = \fr{T_0}{\sqrt{1-E S/2D}}.
\ee
As the temperature decreases, a second minimum appears. The temperature $T_1$ is the temperature at which the potential begins to appear a second minimum. $T_1$ is calculated as follows
\be \label{T1}
T_1=\fr{T_0}{\sqrt{1-9E^2/8D\la _{T_c}}}.
\ee
\hspace{1.25cm}

The strength of EWPT ($S$) is defined as follows
\be \label{S}
S=\fr{v_{c}}{T_c}=\fr{2E}{\la _{T_c}}.
\ee

The one-loop effective potential with the contribution of the Goldstone boson must be written in any gauge, this may be referred to in Appendix F in Ref.~\cite{1101.4665}. Also according to Ref.~\cite{1101.4665}, both EWPT and sphaleron are gauge-independent, Ref. \cite{1101.4665} has carefully demonstrated this with the Nielsen identities order-by-order in perturbation theory. Therefore, we may not consider the one-loop effective potential in any gauge and if we remove the gauges, the contributions of the Goldstone bosons can also be neglected due to their tiny masses.

In the one loop potential analyses, when the one-loop contributions of $h^{\pm}$ and $k^{\pm}$ are added, the parameters in Eq.~(\ref{D-T-E-lamda}) will have contributions from these particles. This is different from the SM.
	
In Eq.~(\ref{D-T-E-lamda}), $E$ has the contributions from $m_{h^\pm}$ and $m_{k^{\pm\pm}}$. This makes $E$ greater than that in the SM. $E$ is the coefficient before $v^3$, the larger the value of $E$, the greater the potential barrier between the two minima of the effective potential, so the phase transition is violently.
	
Similarly, in Eq.~(\ref{D-T-E-lamda}), $\lambda_T$ has the contributions from $m_{h^\pm}$ and $m_{k^{\pm\pm}}$ and let $\lambda_T>0$, the neutral Higgs boson  mass  can reach $125$ GeV.

Next, taking into account daisy loops,  the effective potential will have the form:
\begin{align}
V^{daisy}_{eff}=V_{eff}(v)-\frac{T}{12\pi}\sum_{i=h,W,Z, h^\pm,k^{\pm\pm}}g_i \left \{\left[\frac{m_i^2(v_0)
	v^2}{v^2_0}+\Pi_i(T)\right]^{3/2}-\frac{m_i^3(v_0) v^3}{v^3_0}\right \},\label{daisyloop}
\end{align}
in which the second component on the right hand side of Eq.~(\ref{daisyloop}) is the contribution from daisy loops \cite{carrington,curtin,katz}. Here,  degrees of freedom are given by: $g_Z=3, g_W=6, g_h=1, g_{h^\pm, k^{\pm\pm}}=2$ and
\begin{align}
	&\Pi_W(T)=\frac{22}{3}\frac{m_W^2(v_0)}{v^2_0}T^2, \nonumber\\
	&\Pi_Z(T)=\frac{22}{3}\frac{(m_Z^2(v_0)-m_W^2(v_0))}{v^2_0}T^2,\nonumber\\
	&\Pi_h(T)=\frac{2m_W^2(v_0)+m_Z^2(v_0)+m_h^2(v_0)+2m_t^2(v_0)+2 m_{h^\pm}^2(v_0)+2 m_{k^{\pm\pm}}^2(v_0)}{4v^2_0}T^2. \label{2.164}
\end{align}

Following steps in in Refs.~\cite{carrington,curtin,katz}, we calculate the contributions of the polarization tensors of $h^\pm, k^{\pm\pm}$, the results are as follows:

\begin{align}
	&\Pi_{h^\pm}(T)=\frac{1}{6}\frac{m_{h^\pm}^2(v_0)}{v^2_0}T^2, \crn
	&\Pi_{k^{\pm\pm}}(T)=\frac{1}{6}\frac{m_{k^{\pm\pm}}^2(v_0)}{v^2_0}T^2\,.
\end{align}

In Ref.~\cite{carrington}, when $m/T \sim 1$ we can skip the daisy loop component. Since these masses of charged Higgs bosons are large (see in the Table \ref{giams}), the critical temperature is around $120$ GeV, and $S=v_c/T_c>1$; thus $m(v_c)/T_c\sim 1$. Therefore, we do not take into account the daisy loops of the charged Higgs bosons. However, we also calculate the contributions of the daisy loops of $h^\pm, k^{\pm\pm}$ for comparisons between cases as shown in Fig.~\ref{ssdb}.

However, in the temperature region higher than $T_C\sim 120$ GeV, these daisy loops will make an important contribution. For example the calculations in Ref.~\cite{espinosa}, the authors calculated the effective potential in the case of very high temperatures, around $500$ GeV (see in Fig. 1), so $m/T$ will be very small. The conditions of $m/T$ (or $\alpha$, $\beta$) in Ref.~\cite{espinosa} are the same as in Ref.~\cite{carrington}.

We randomly select the masses of $h_{\pm}$ and $k_{\pm\pm}$, then recalculate $S$ for both cases, namely: with and without daisy loops. Details are as shown in Table \ref{giams}.
	\begin{table}[!ht]
	\centering
	\caption{Strength $S$ has  daisy loops without $h^\pm, k^{\pm\pm}$ or not}
	\begin{tabular}{cc|c|c|c|c|c|c|c}\hline\hline
		$m_{k^{\pm\pm}}$ [GeV]&$m_{h^{\pm}}$ [GeV]&$V_{C-nodaisy}$&$T_{C-nodaisy}$&$S_{nodaisy}$&$V_{C-daisy}$&$T_{C-daisy}$&$S_{daisy}$&$S_{daisy}/S_{nodaisy}$\\ \hline
		230 & 286.1385 & 208 & 119.2& 1.745 & 208 & 124.9 & 1.665 & 0.95\\
		240 & 280.3134 & 209 & 119.1 & 1.755 & 209 & 124.8 & 1.675 & 0.95\\
		250 & 120 & 134 & 127.9 & 1.048 & 126 & 135.5 & 0.929 & 0.89\\
		258 & 140 & 145 & 125.8 & 1.153 & 140 & 132.9 & 1.053 & 0.91\\
		264 & 170 & 160 & 123.4 & 1.297 & 155 & 130 & 1.192 & 0.92\\	
		\hline\hline
	\end{tabular}\label{giams}
\end{table}

The ring-loops or daisy-loops reduce the strength of the phase transition by $2/3$ times (we also talked about this in the Introduction section). This is covered in detail in Ref.~\cite{r23, 1101.4665}. However, from Table \ref{giams}, it follows that daisy loops (without daisy loops of $h^\pm, k^{\pm\pm}$) reduce $S$ by $0.89-0.95$ times {\it (but not} $2/3$). If taking into account the daisy loops of $h^\pm$ and $k^{\pm\pm}$, $S$ will reduce by $0.78-0.9$ times.

Therefore, according to Ref. \cite{zb}, the largest strength is about $2.4$ in the Landau gauge. However,  if we add daisy-loops, the largest $S$ is only about $2.28$ ($2.4\times 0.95$), when $m_{h^{\pm}}$ and $m_{k^{\pm\pm}}$ are in the range from  $0$ to $300\, \mathrm{ GeV}$. This is also one reason for our argument about $S$ cannot be too large.

For the perturbation theory, the parameters in the Higgs potential should always be less than $4\pi$ ~\cite{garcia}. However, $\mu_{hk}$ has a dimensions of mass, with radiation corrections, to avoid breaking charge conservation, $\mu_{hk}\ll 4\pi. max(M_h,M_k)$ ~\cite{garcia}. For a large $\mu_{hk}$ case, the charge breaking minimum is no a global minimum Ref.~\cite{garcia}. When calculating the first order EWPT, $m_{h^\pm/k^{\pm\pm}}<300$ GeV, so we obtain

\be
	q.v_0<300; \hs p.v_0 < 300,
\ee
Thus, $p,q<1.219<4\pi$.

Let us discuss  the case of $0<m_{h^\pm/k^{\pm\pm}}<300\, \mathrm{ GeV}$. This is just the range of possible masses by numerical solution. From this mass domain we think that there is a case where smallness $m_{h^{\pm}}$ and $m_{k^{\pm\pm}}$ (a weakly coupling) also leads to a first order EWPT. We assume that non-perturbative corrections make significant contributions and do not change the results. However, this is not the case. Because $m_{h^\pm}$ is small, $m_{k^{\pm\pm}}$ must be large enough to enable a first order EWPT and vice versa. So the above mass domain is only a general record.

Also the case of smallness $m_{h^\pm}$ or $m_{k^{\pm\pm}}$ would be difficult because their signals are not found in the experimental data. One only see the signals of the SM particles. 

We also note that this model has only one mass generation particle (Higgs), so when $m_h$ and $m_k$ are too large, $T^2_0$ in Eq.~(\ref{D-T-E-lamda}) would be have negative values. Therefore the zero temperature potential does not break electroweak  symmetry because the Higgs mass squared received large (positive) loop contributions from the new heavy bosons.

So it can be seen from this model that we want a model with a larger the strength of phase transition, in addition to having many triggers,
the model must also need more than one SM-like Higgs particle.

\section{Sphaleron in the Zee-Babu model and proving the sphaleron scaling law}\label{sec3}

In a chiral hypothesis, for example, the electroweak theory there is another fascinating sort of progress called topological transition, which is a transition progress between vacua. At the point when such a transition happens, the baryon number will be violated  and it opens up the opportunities for the electroweak hypothesis to fulfill the main condition of Sakharov. In this transition, there are two types: Instanton and Sphaleron.

However, with a first order EWPT at high temperature the sphaleron process is stronger than instanton. The universe is accelerated while this phase transition occurs thought out sphaleron, so the condition for the departure from thermal equilibrium (ie., EWPT) is $\Ga_{sph} \ll H_{rad}$ or the decoupling condition.

As mentioned in the introduction, there are two ways to calculate the sphaleron energy. We will do this calculation using the second method suggested by Manton and Klinkhamer \cite{10}. The ansatz approaches  with profiles functions in Ref.~\cite{10} to approximate a sphaleron solution without solving the  EOMs.

\subsection{The zero temperature sphaleron at tree level}

As a first step, we have to calculate the sphaleron energy at 0K, which is the maximum boundary value and
as well at the same time to test the convergence of this functional.

In the ZB model, the sphaleron energy functional
consists of three main components: The first component is gauge bosons, the second and third components are kinetic and potential of Higgs scalars, respectively,
\be
E_{sph}=\int dx^3 \left[ \fr{1}{4}W_{ij}^a W_{ij}^a +(D_i \phi)^\dagger (D_i \phi)+ V(\phi,h^{\pm},k^{\pm\pm})_{tree}\right].
\label{e1}
\ee

We ignore the contributions of the hypercharge gauge field because they are very small \cite{10}. We also assume the static field approximation (ie., $W^a_0=0$) and sphaleron has a spherically symmetric form \cite{11}:
\be
\left\{ \begin{array}{ll}
\phi(r)=\fr{v_0}{\sqrt{2}} h(r) i n_a \sigma \left(\begin{array}{cc} 0\\1\end{array} \right),\\
W_i ^a (r)=\fr{2}{g}\ep ^{aij} n_j \fr{f(r)}{r},
\end{array}\right.
\label{ham}
\ee
where $n_i \equiv \fr{x^i}{r}$ and $r$ is the radial coordinate in the spherical coordinates. Here, $h(r)$ and $f(r)$ have boundary conditions
\be
\left\{ \begin{array}{ll}
	h(r\to 0)=f(r\to 0)=0,\\
	h(r\to \infty)=f(r\to \infty)=1.
\end{array}\right.
\label{dk}
\ee

There are two types of  profile ansatzs: the ansatz with scale-free parameters firstly introduced in Ref.~\cite{10}, and the smooth ansatz as in Refs.~\cite{11,gauge0}. In addition, with composite Higgs, models can have other profiles \cite{phu2spha}. According to the calculation in Ref.~\cite{plk}, the ansatz with scale-free parameters is more accurate. So we use the ansatz with two scale-free parameters $a$ and $b$:
\be
f(\xi)=  \left\{\begin{array}{c} f_1(\xi)=\fr{\xi^2}{2a^2} , \xi \le a\\
	f_2(\xi)=1-\fr{a^2}{\xi^2} ,  \xi \geq a
	
\end{array} \right.
\label{dk1}
\ee
\be
h(\xi)=  \left\{\begin{array}{c}h_1(\xi)=\fr{4\xi}{5b} , \xi \le b\\
	h_2(\xi)=1-\fr{b^4}{5 \xi^4} ,  \xi \geq b
\end{array} \right.\,,
\label{dk2}
\ee
where \bea
\xi\equiv gv_0r\rightarrow d\xi =gv_0dr.
\eea

The potential at tree level can be rewritten as follows:
\bea
V(\phi,h^\pm,k^{\pm\pm},T=0)_{tree}&=&\mu^2\phi^{\dagger}\phi +\la  (\phi^{\dagger}\phi)^2
\crn
&+&2p^2{\vert{h}\vert}^{2} \phi^{\dagger}\phi+2q^2{\vert{k}\vert}^{2} \phi^{\dagger}\phi+\mathcal{V}\,
\label{2aa}
\eea
where
\bea
\mathcal{V}&=&u^2_1{\vert{h}\vert}^2+u_2^2{\vert{k}\vert}^2+\la _h{ \vert{h}\vert}^4 + \la _k{ \vert{k}\vert}^4 +\la _{hk}{\vert{h}\vert}^2{ \vert{k}\vert}^2
\crn
&+&(\mu_{hk} h^{2} k^{++}+ H.c).
\label{2a}
\eea
We next accept \textit{an approximation}:
\bea
h^{\pm}(r)&=& c_1,\\
k^{\pm\pm}(r)& = &c_2,
\eea
where \textit{$c_1$ and $c_2$ are very small}. So $\mathcal{V}$ in Eq. (\ref{2a}) becomes constant which can be set to $\La $.

We are able to give this approximation because the mass-generating process for particles is their interactions with the Higgs field. $\mathcal{V}$ contains components that do not interact with the Higgs boson. Components in $\mathcal{V}$ as a mechanism to explain the tiny mass of neutrino problem \cite{zeebabu, zeebabub}. The coefficients in $\mathcal{V}$ are undefined. The values of these coupling constants need to be determined through decay channels.

After putting the ansatzs into Eq. (\ref{2aa}), we will get
\be
V(\phi,h^\pm,k^{\pm\pm},T=0)_{tree} = -\fr{\mu'^2v_0^2}{2}h^2+\fr{\la  v_0^4}{4}h^4+\La\label{29}
\ee
where $\mu'^2=\mu^2+4p^2c^2_1+4q^2c^2_2=constant\approx \mu^2$. We can minimize the above potential and eliminate $\La $, then obtain
\be
V(\phi,h^\pm,k^{\pm\pm},T=0)_{tree}=\fr{\la  v_0^4}{4}(h^2-1)^2.
\label{2bb}
\ee

There are two conditions to test the process of calculating sphaleron energy. The first is the scaling law. The second condition is the decoupling condition in EWPT. As indicated in next sections, both of these conditions are satisfied. If their bubbles are too large (i.e., $c_1$ and $c_2$ are too large), $\mu'$ is very larger than $\mu$ and Eqs.~(\ref{2bb}),(\ref{sph,0}), (\ref{sphT}) have additional contributions from $c_1$ and $c_2$. Therefore, if $c_1$ and $c_2$ are large, the sphaleron energy will be large or they can break the scaling law.

We accept the scenario where $c_1$ and $c_2$ are small. If $c_1$ and $c_2$ are large, the sphaleron calculations need to be recalculated, especially the normalization processes in Eq.~\eqref{29}, so the integrals \eqref{sph1} and \eqref{sphT} depend on the magnitude of $c_1$ and $c_2$.

In addition, when the sphaleron energy is too large, $\beta/H$ will be large, resulting in a small gravitational wave generation, as shown in Fig.~\ref{betaalpha}.

Thus, the approximation of the contribution of $\mathcal{V}$ and the smallness $c_1$ and $c_2$ are acceptable, considering the possibility of generating the largest gravitational waves and satisfying the above two conditions.

After making a few simple transformations, the zero temperature sphaleron energy at tree-level is
\be
E_{sph,tree}^{0K}=\fr{4 \pi v_0}{g} \int_0^\infty d\xi \left[ 4\left(\fr{df}{\xi}\right)^2
+\fr{8 f^2(1-f)^2}{\xi^2}  +\fr{\xi^2}{2}\left(\fr{dh}{\xi}\right)^2
+ h^2(1-f)^2 +\fr{\xi^2}{g^2}\fr{\la  }{4}(h^2-1)^2\right]
\label{sph,0}
\ee

Expanding Eq.(\ref{sph,0}) by using Eqs. (\ref{dk1}) and (\ref{dk2}), we get
\bea
E_{sph,tree}^{0K}&=&\fr{4 \pi v_0}{g} \int_0^a d\xi \left[ 4\left(\fr{df_1(\xi)}{\xi}\right)^2
+\fr{8 f^2_1(\xi)(1-f_1(\xi))^2}{\xi^2}  \right] \crn
&+&\fr{4 \pi v_0}{g} \int_0^a d\xi \left[ \fr{\xi^2}{2}\left(\fr{df_1(\xi)}{\xi}\right)^2
+\fr{8 h^2_1(\xi)(1-f_1(\xi))^2}{\xi^2} +\fr{\xi^2\la }{4g^2} \left(h_1^2(\xi)-1\right)^2\right] \crn
&+& \fr{4 \pi v_0}{g} \int_a^b d\xi \left[ 4\left(\fr{df_2(\xi)}{\xi}\right)^2
+\fr{8 f^2_2(\xi)(1-f_2(\xi))^2}{\xi^2}  \right]\crn
&+&\fr{4 \pi v_0}{g} \int_a^b d\xi \left[ \fr{\xi^2}{2}\left(\fr{dh_1(\xi)}{\xi}\right)^2
+\fr{8 h^2_1(\xi)(1-f_2(\xi))^2}{\xi^2} +\fr{\xi^2\la }{4g^2} \left(h_1^2(\xi)-1\right)^2\right] \crn
&+& \fr{4 \pi v_0}{g} \int_b^\infty d\xi \left[ 4\left(\fr{dh_2(\xi)}{\xi}\right)^2
+\fr{8 h^2_2(\xi)(1-h_2^2(\xi))}{\xi^2}  \right]\crn
&+&\fr{4 \pi v_0}{g} \int_b^\infty d\xi \left[ \fr{\xi^2}{2}\left(\fr{dh_2(\xi)}{\xi}\right)^2
+\fr{8 h^2_2(\xi)(1-f_2(\xi))^2}{\xi^2} +\fr{\xi^2\la }{4g^2} \left(h_2^2(\xi)-1\right)^2\right],
\label{sph1}
\eea
where $g^2=\fr{G_F 8 m^2_W}{\sqrt{2}}$; $G_F=1.166 \times 10^-{15} \, \textrm{GeV}^{-2}$; $m_W=80.39\, \textrm{GeV}$; $\la  \simeq 0.12$.

To minimize this energy according to the parameters $a$ and $b$, we get the values of sphaleron energy given in  Table \ref{sphtree}.
\begin{table}[!ht]
	\centering
	\caption{The zero temperature sphaleron energy at tree level}
	\begin{tabular}{ccccc}\hline\hline
	$a$&$b$&$v_0$[\textrm{GeV}]&$E_{sph}^{T=0K}[\textrm{GeV}]$&$E_{sph}^{T=0K}/v_0$\\ \hline
$2.4$&$1.28$&246&$10245.7$&$41.649$\\ 	
		 \hline\hline
	\end{tabular}\label{sphtree}
\end{table}

Note that Eq.~(\ref{sph1}) has a symmetry between $a$ and $b$. If we assume $a>b$ or $b>a$, the above integral gives the same minimum value. The maximum sphaleron energy which is one at 0K, can be estimated at around $10.2457$ TeV.

\subsection{Sphaleron at the non-zero temperature and the scaling law}
\subsubsection{Sphaleron at the non-zero temperature}\label{snzero}

We just need to replace $V(\phi,h^{\pm},k^{\pm\pm})$ by $V_{eff}$ in the sphaleron energy at 0K, then we will get sphaleron energy at the non-zero temperature. However, we first use the effective potential without daisy loops, which we will explain later. The sphaleron energy at the non-zero temperature is rewritten as follows
\be
E_{sph}^{T}=\fr{4 \pi v_0}{g} \int_0^\infty d\xi \left[ 4\left(\fr{df}{\xi}\right)^2
+\fr{8 f^2(1-f)^2}{\xi^2}  +\fr{\xi^2}{2}\left(\fr{dh}{\xi}\right)^2
+ h^2(1-f)^2 +\fr{\xi^2}{g^2 v^4_0}V_{eff}(h,T)\right],
\label{sphT}
\ee
where
\be
V_{eff}(h,T)=D(T^2-T^2_0)v^2-ET|v|^3+\fr{\la _T}{4}v^4.
\label{VT}
\ee
Notice that
\be
\phi^\dagger \phi = \fr{v_0^2 h^2(r)}{2}\,.
\label{hh}
\ee

On the other hand, ignoring the contribution of the Goldstone bosons, the Higgs doublet becomes
\be
\phi=\fr{1}{\sqrt{2}} \left(\begin{array}{c}
	0 \\
	v
\end{array}\right)\to \phi^\dagger \phi=\fr{v^2}{2}.
\label{h1}
\ee
Form Eq.~(\ref{hh}) and Eq.~(\ref{h1}), we obtain
\be
v=v_0 \times h(r)
\label{hp}
\ee
Inserting Eq.~(\ref{hp}) into Eq.~(\ref{VT}) we then get
\bea
V_{eff}^{T}=D(T^2-T^2_0)v_0^2 h^2-ET|h|^3 v_0^3+\fr{\la _T}{4}v_0^4h^4.
\label{ft}
\eea

We must rescale the variable $\xi$ and the radial function $h(\xi)$ as follows:
\[
\tilde{\xi}=\fr{\xi}{s}; \hspace{1cm}
\tilde{h}=sh;\hspace{1cm} \text{with }s\equiv\fr{v(T)}{v_0}.
\]

then Eq. (\ref{ft}) becomes
\bea
V_{eff}(h,T)&=&\fr{\la_T}{4}v^4-ETv^3+D(T^2-T_0^2)v^2\nn\\
&=& \fr{\la_T v_0^4}{4}s^4h^4-ETv_0^3s^3h^3+D(T^2-T_0^2)v_0^2s^2h^2\nn\\
&=&\fr{\la_T v_0^4}{4}s^4(h^4-1)-ETv_0^3s^3(h^3-1)+D(T^2-T_0^2)v_0^2s^2(h^2-1)\nn\\%
&+&\mathcal{C}(T)\label{ct}.
\eea

In Eq.~(\ref{ct}), $\mathcal{C}(T)$ will be removed by the normalized conditions of the effective potential, so that $E_{sph}^{T}$ does not diverge. The sphaleron energy will be rewritten as
\bea
E_{sph}^T&=&\fr{4\pi v_0}{gs}\int_0^\infty d\xi \Bigg\{s^2\Bigg[4\left(\fr{df}{d\xi}\right)^2+\fr{8}{\xi^2}f^2(1-f)^2+\fr{\xi^2}{2}\left(\fr{dh}{d\xi}\right)^2+h^2(1-f)^2\Bigg]\crn
&+&\fr{\xi^2}{g^2s^2}\Bigg[\fr{\la_T}{4}s^4(h^4-1)-\fr{ET}{v_0} s^3(h^3-1)+\fr{D(T^2-T_0^2)}{v_0^2} s^2(h^2-1)\Bigg\}\label{sphat}.
\eea

In Eq.~(\ref{sphat}) we can separate out the sphaleron energy of the gauge fields in terms of the variable $\xi$ as follows \cite{plk}
\be
\begin{aligned}
	E^T_{gauge}&=\int \fr{1}{4}W_{ij}^a W_{ij}^a dx^3=\fr{4\pi v}{g}\int_0^\infty d\xi\left[4\left(\fr{df}{d\xi}\right)^2+\fr{8f^2(1-f)^2}{\xi^2}\right].
\end{aligned}
\label{mar5}
\ee

With different $S$ values, we get the specific mass of $h^{\pm}$ and $k^{\pm\pm}$. Then substitute them in Eq.~(\ref{sphat}), and minimize the energy functional to find the values $a,b$. Finally,  substituting  $a,b$ which we just found with the mass of the particles in Eq.~(\ref{sphat}), then we get the sphaleron energy at any temperature in the range below $T_c$.

To ensure that the  baryon number remains unchanged during the expansion of the universe, we need the baryon erosion avoidance condition, which is also called the de-coupling condition or the third Sakharov condition. During the phase transition (that is, when the temperature drops from $ T_c $ to $ T_0 $), the rate of violation of the baryon number must be less than the Hubble rate. Moreover, EWPT occurs after inflation and belongs to the period dominated by radiation, so the Hubble rate is $H^2_{rad}=\fr{4\pi^3g_*}{45 m_{pl}^2}T^4$. Thus $\Ga _{sph} \ll H_ {rad}$ \cite{spha-huble,decoupling, decouplingb, decouplingc, decouplingd,11} as the result:
\be\label{decoupling_equation}
\fr{E_{sph}^T}{T}-7ln\left(\fr{v(T)}{T}\right)+ln\left(\fr{T}{100\,  \textrm{GeV}}\right)>(35.9-42.8).
\ee

For detailed illustration, we calculate the sphaleron energy with $S=1.2$, the results obtained are as shown in Table \ref{decoupling_table}. These results satisfy the decoupling and supercooling conditions \cite{Fuyuto,Fuyutob,Fuyutoc}.

\begin{table}[!ht]
	\centering
	\caption{The sphaleron energy at $S=1.2$, $m_{k^{\pm\pm}}=256.169 \,  \textrm{GeV}$, $m_{h^{\pm}}=218.500 \,  \textrm{GeV}$.}
	\begin{tabular}{cccc|cccc|c}\hline\hline
		$T[\textrm{GeV}]$&$v [\textrm{GeV}]$&$a$&$b$ &$E^T_{gauge}$&$E^T_{V}$& $E_{sph}^T [\textrm{GeV}]$&$E^T_{gauge}/E_{sph}^T$&De-coupling: Eq.~(\ref{decoupling_equation})\\ \hline	
		$T_c=	118.104$&141.724&2.591&2.855&2792.88&170.15&5245.45&53.24\%&$43.304 > 42.8$
		\\ \hline
		$T_{n1}=117.000$ &150.056&2.540&2.715&3016.16&209.74&5612.83&53.73\%&$46.38>42.8$
		\\ \hline
		$T_{n2}=	115.000$&161.346&2.490&2.579&3307.95&256.40&6103.09&54.20\%&$50.83>42.8$
		\\ \hline
		$T_{n3}=	113.000$&170.055&2.461&2.499&3528.46&289.40&6478.11&54.46\%&$54.58>42.8$
		\\ \hline
		$T_{n4}=	111.000$&177.238&2.440&2.444&3708.61&315.44&6786.33&54.64\%&$57.96>42.8$
		\\ \hline
		$T_{0}=	105.794$&191.66&2.420&2.425&4042.93&401.27&7407.11&54.58\%&$65.91>42.8$
		\\ \hline\hline
	\end{tabular}\label{decoupling_table}
\end{table}	

We see that $E_{sph}^T$ at any temperatures satisfy this condition, see in the eighth column of the Table \ref{decoupling_table}, the gauge component contributed the largest, accounting for about 55\%. Although the results are only examined for one value of $S$, the laws of sphaleron energy hold for other $S$.

Furthermore, according to Table \ref{decoupling_table}, the potential contribution ($E^T_V$) is always less than $5.50\%~(\sim 401.27/7407.11)$. Therefore, when adding daisy loops, the contribution from potential to the sphaleron energy becomes smaller or actually the contribution of daisy loops is too small compared to the others. Because of Eq.~(\ref{daisyloop}), $V^{daisy}_{eff}<V_{eff}(v)$. Therefore, to calculate the sphaleron energy, we just use the effective potential without daisy loops.

\begin{figure}[h!]
	\centering
	\begin{subfigure}{.5\textwidth}
		\centering
		\includegraphics[width=1\linewidth]{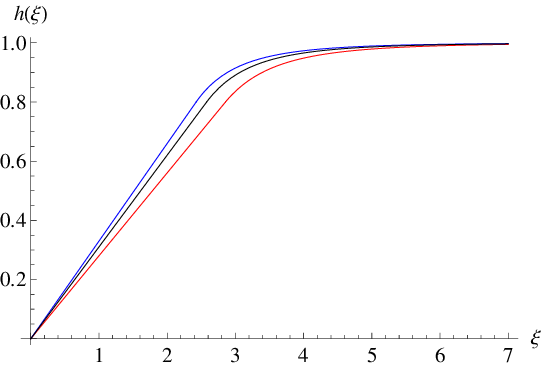}
		\caption{The $h(\xi)$ profile with $S=1.2$. The red line: $T=T_c$, the blue and black line: $T=T_0$ and $T_{n2}$.}
		\label{pf1}
	\end{subfigure}%
	\begin{subfigure}{.5\textwidth}
		\centering
		\includegraphics[width=1\linewidth]{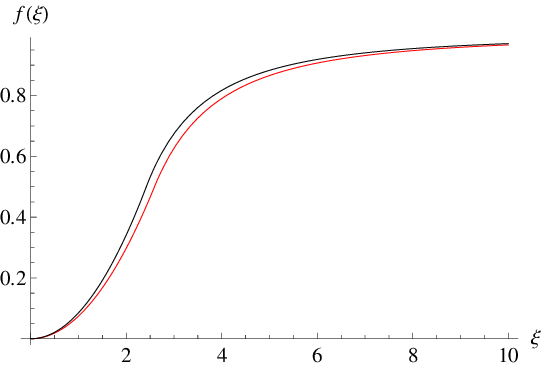}
		\caption{The $f(\xi)$ profile  with $S=1.2$. The red line: $T=T_c$, the black line: $T=T_{0}$.}
		\label{pf2}
	\end{subfigure}
	\caption{Nuclear bubbles}
	\label{pf0}
\end{figure}

The profile functions of $h(\xi)$ and $f(\xi)$ are plotted in Fig.~\ref{pf0}. These functions rapidly advance to $1$  as $\xi$ increases. As $\xi$ increases, the faster the functions approach $1$, the thinner the wall of bubbles.

\begin{figure}[h!]
	\centering	
	\includegraphics[width=0.6\linewidth]{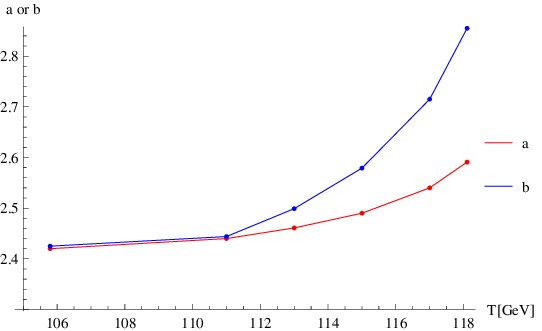}
	\caption{The values of a and b as functions  of $T$}
	\label{ab}
\end{figure}

According to the results of Table \ref{decoupling_table}, the values of $a,b$ decrease as the temperature decreases. When the temperature is near $T_c$, $a$ and $b$ decrease very quickly (see  Fig. \ref{ab}).  $a,b$ are two free parameters, but they are specific to the thickness of the wall of bubbles. This leads to an initial prediction that the radius of bubbles will increase when the temperature goes down because the thickness of wall of bubbles decrease. We will verify this again in the next section after calculating $\beta/H^*$.

\subsubsection{The scaling law}

\textit{The sphaleron scaling law} \cite{47, Ahriche1, Ahriche2,Ahriche2b}: The sphaleron energy at a temperature below the critical temperature is linearly proportional to VEV at that temperature
\be
E_{scaling}^T\approx \fr{v(T)}{v_0}E_{sph}^{T=0K}.
\label{mar8}
\ee
This law is only an approximate expansion. But it has a necessary meaning in order to double-check the calculation results.

\begin{figure}[h!]
		\centering
		\includegraphics[width=0.6\linewidth]{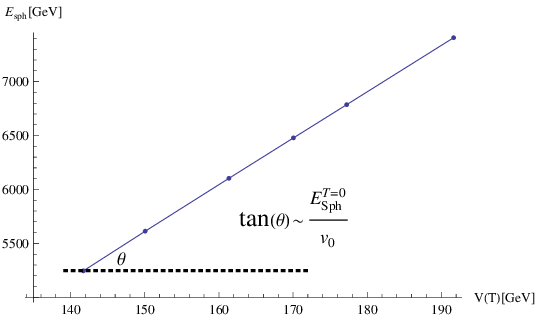}
		\caption{The sphaleron scaling law}
		\label{scaling}	
\end{figure}

Using data in Table \ref{decoupling_table}, we have plotted $E_{sph}^T$ (in the seventh column) as a function of $v(T)$ (the second column) in Fig. \ref{scaling}.
 From Fig.~\ref{scaling}, $E^T_{sph}$ increases linearly with $v(T)$ and the slope of this line has the typical range $37-38$. But according to Eq.~(\ref{mar8}), this slope should be equal to $\fr{E_{sph}^{0K}}{v_0}$, so that $\tan(\theta)=\fr{E_{sph}^{0K}}{v_0}$. From the Table \ref{sphtree}, we get $\tan(\theta)=\fr{E_{sph}^{0K}}{v_0}=\fr{10245.7}{246}=41.649$. We therefore find that the above law is on average 10\% more than it actually is.

Hence, we revise the sphaleron scaling law as follows:

\be
E_{scaling}^T\approx v(T).\fr{0.9E_{sph}^{0K}}{v_0}.
\label{mar9}
\ee
This can be clearly seen when comparing Eq.~(\ref{sph1}) with Eq.~(\ref{sphT});
and $V(\phi,h^{\pm},k^{\pm\pm})$ is very different from $V_{eff}(h,T)$. However, this correction is not large because the contribution of the effective potential component is small, which is also mentioned in Ref.~\cite{Ahriche1}. In addition this correction is also consistent with the results given in Ref.~\cite{plk}.

\section{GW production from EWPT in the Zee-Babu model}\label{sec4}

Electroweak phase transition with nuclear bubbles that has
 spherical symmetry. As the temperature drops from the critical temperature, the collisions of bubbles destroy the spherical symmetry. This produces gravitational radiation. The high velocity and energy density of the bubbles will provide the necessary conditions for the generation of gravitational radiation. There are two sources of gravitational waves: the direct collision of the bubbles and the chaos in the plasma state due to the motion of the bubbles \cite{2gw, 2gwa,2gwb,7gw,7gwb}.

\subsection{The parameters $\beta/H^*$ and $\al $}

Two important parameters for calculating gravitational waves that have been derived from many past studies \cite{phu1gw,phu2gw,phu2gwb,phu3gw,phu4gw,phu5gw,phu7gw,phu8gw,1gw,2gw,2gwa,2gwb,4gw,7gw,7gwb,8gw,9gw} are latent heat ($\al $) and $\beta/H^*$.

The first parameter, $\beta$ is often called the inverse of the time of the phase transition. This quantity determines the size of the bubble at the time of collision, so it is characteristic of the frequency at which the GW signal is largest \cite{1gw}.

\subsubsection{The transition rate}

The sphaleron rate below the critical temperature is usually defined as:
\be
\Ga (t)=A(t) e^{-S(t)}.
\label{gma1}
\ee
We deduce the derivative of this rate with respect to time:
\be
\fr{d \Ga (t)}{dt}=\fr{d A(t)}{dt} e^{-S(t)}-A(t)\fr{d S(t)}{dt} e^{-S(t)}.
\label{gmaa}
\ee
in which, $A(t)\sim T^4$ so $\fr{d A(t)}{dt}\approx\fr{d(T^4)}{dt}= \fr{d(T^4)}{dT}\fr{dT}{dt}=4T^3 \fr{dT}{dt}$ and $S(t)$ is the Euclidean action of a critical bubble.

The change of temperature over time, calculated through the Hubble coefficient $H$ gives $\fr{dT}{dt}=-T \times H$. So Eq.~(\ref{gma1}) becomes:
\bea
\fr{d \Ga (t)}{dt}=-4 T^4H e^{-S(t)}+T^5H \fr{d\left(S(t)\right)}{dT},
\eea

\subsubsection{The parameter $\beta/H^*$}

The parameter $\beta$ is defined as follows:
\bea
\beta&\approx&  \fr{1}{\Ga }  \fr{d \Ga (t)}{dt}=\left[-4 H+TH \fr{d\left( S(t)\right)}{dT}\right],\label{4h}
\eea
with $H\approx3.36\times10^{-14}$. So we can ignore the term $-4H$ in Eq.~(\ref{4h}) so it can be rewritten as follows
\bea
\fr{\beta}{H^*}&\approx& \left[T \fr{d\left( S(t)\right)}{dT}\right]_{T=T_C}.
\label{beta1}
\eea

We can approximate $S(t)\approx S_3(T)/T$. Here $S_3(T)$ is the action for the $O(3)$ symmetric bounce action showing transitions between vacua of the Higgs potential, excluding gauge fields. Therefore $S_3(T)$ can be approximated by sphaleron energy without the gauge component. From  the values in Table \ref{decoupling_table}, it follows that the gauge contribution accounts for 55\% of the sphaleron energy. Therefore, a possible approximation is	
	\be
	\fr{\beta}{H^*}\approx \left[T \fr{d\left( S(t)\right)}{dT}\right]_{T=T_C}\approx\left[T \fr{d\left( S_3/T\right)}{dT}\right]_{T=T_C}\approx \left[T \fr{d\left(0.45\fr{E^T_{sph}}{T}\right)}{dT}\right]_{T=T_C}.
	\label{beta}
	\ee

Here, $\fr{\beta}{H^*}$ is dimensionless and depends mainly on the sphaleron energy and the shape of the potential at the time of nucleation. $H^*$ is defined  as follows \cite{8gw}
\bea
H^*&=& 4.5 \times 10^{-22}\left[\fr{T_c}{MeV}\right]^2\,\textrm{ MeV} \left[\fr{g_*}{10.75}\right]^2;\,  g^*=106.75.
\label{hsao}
\eea

In addition we have three quantities that are related to the nuclear bubble defined via the $\beta$ parameter,
\bea
\begin{cases}
	\beta^{-1}\Ga (t): \text{the number of nuclear bubbles,}\\
	R_{b}\sim v_{b} \beta^{-1}: \text{the radius of nuclear bubble,}\\
	\beta^{-1}: \text{the typical time-scale of transition process.}
\end{cases}\label{radiusvb}
\eea
in which, $v_b$ is the velocity of the wall of bubble.

The parameter $\beta$ is difficult to determine, because we need to know the change of the sphaleron energy functional around the critical temperature. But Eq.~(\ref{sphT})  only tells us the energy at $T\leq T_c$. And the value of the sphaleron energy when $T>T_c$ is often difficult to determine, because the sphaleron rate in the thermal equilibrium is very small due to the sphaleron processes can be wash-outed by the inverse process.

From Eq. (\ref{beta})
it is reasonable  the following approximation:
	\be
	\beta/H^*\approx \bigg[\fr{0.45 E^T_{sph}}{T}\bigg]_{T=T_c}.\label{betah}
	\ee
	
In Sec. \ref{snzero}, we can omit daisy loops in the sphaleron energy. Thus daisy loops also do not contribute to the value of $\beta/H$.

\subsubsection{The parameter $\al $}

The second important parameter characterizing the spectrum of gravitational waves is $\al $ that is the ratio of latent heat ($\ep $) released in the radiation energy density ($\rho^*_{rad}$). $\al $ is determined at the nucleation temperature  as follows \cite{2gw, 2gwa,2gwb}:
\begin{align}
\al &=\fr{\ep }{\rho_{rad}^*},\\
\ep &=\left(V_{eff}(v(T),T)-T \fr{d}{d T}V_{eff}(v(T),T)\right)_{T=T_C},\label{alla}\\
\rho_{rad}^*&=g^* \pi^2 \fr{T^4_c}{30}=106.75\pi^2 \fr{T^4_c}{30}.
\label{all}
\end{align}

This parameter depends only on the configuration of the effective potential. Thus in all models the parameter which is easily  calculated by usual way when the effective potential is known, determines the strength of the phase transition.

\begin{figure}[h!]
	\centering
	\begin{subfigure}{.5\textwidth}
		\centering
		\includegraphics[width=1\linewidth]{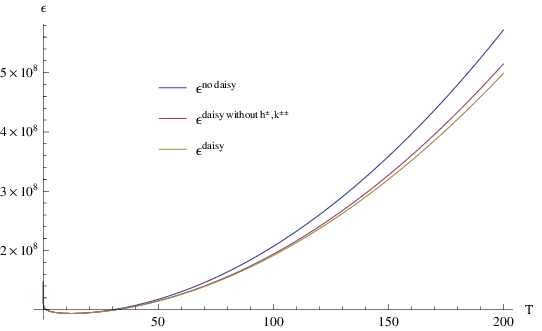}
		\caption{$\epsilon^{no daisy}$ and $\epsilon^{daisy}$ change with temperature.}
		\label{ssda}
	\end{subfigure}%
	\begin{subfigure}{.5\textwidth}
		\centering
		\includegraphics[width=1\linewidth]{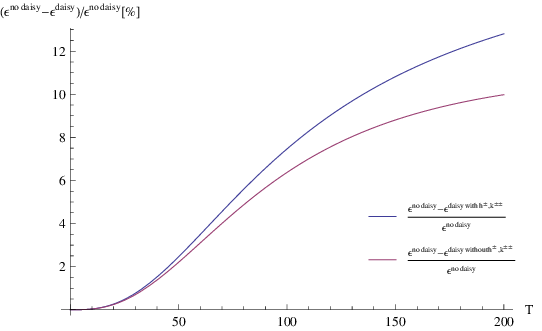}
		\caption{Difference between $\epsilon^{no daisy}$ and $\epsilon^{daisy}$.}
		\label{ssdb}
	\end{subfigure}
	\caption{The latent heat, $m_{h_{\pm}}=120$ GeV and $m_{k_{\pm\pm}}=280$ GeV.}
	\label{ssd}
\end{figure}

Taking into account Eq.~(\ref{alla}), we can use the effective potential with or without daisy loops. The value of latent heat of two cases through Fig.~\ref{ssdb} which is plotted in two definite values of $h_{\pm}$ and $k_{\pm\pm}$.

In Fig.~\ref{ssdb}, there is a difference between $\epsilon^{no daisy}$ and $\epsilon^{daisy}$ but it is always less than $12\%$ (If the daisy loops of $h^\pm$ and $k^\pm$ are not taken into account, the difference is less than 10\%.). 
Looking at Fig.~\ref{ssda}, we see that the higher the temperature, the larger the difference. Therefore, we can use the effective potential without daisy loops to calculate $\epsilon$ or gravitational waves.

\subsection{The GW energy density parameter}

With the effective potential without daisy loops in the above sections, we can compute the quantities $\al$ and $\beta/H^*$ and from that we can predict for the GW spectrum which is characterized by the density parameter.

There are three processes which are involved in the production of GWs at a first-order PT: Collisions of bubble walls, sound waves in the plasma and Magnetohydrodynamic (MHD) turbulence \cite{71}. These three processes generically coexist, and the corresponding contributions to the stochastic GW background should linearly combine, at least approximately, so that \cite{71}	
	\be
	h^2\Om_{GW}=h^2\Om_{Coll}+h^2\Om_{sw}+h^2\Om_{tur}.
	\ee	
	We have three different scenarios depending on the bubble wall velocity \cite{71}. The bubble wall velocity is defined via $\al $:
	\be
	v_b= \fr{\fr{1}{\sqrt{3}}+\sqrt{\al ^2+2\fr{\al }{3}}}{1+\al }.
	\label{al}
	\ee
	
We estimate that $v_b\sim 1$. This will be clearly seen on Table \ref{omegah}. Therefore, the generation of gravitational waves has only two components: the sound wave and turbulence \cite{71, 73, 73a, 73b, 73c, 73d, 73e}.
	
	The gravitational wave energy density parameters of sound waves is defined through $\al $ and $\beta/H^*$ as follows \cite{phu5gw, 71}:
	\be
	h^2\Om_{sw}(f) \simeq 2.65 \times 10^{-6} v_bk^2_w \left[\fr{H_*}{\beta}\right] \left[\fr{\al }{1+\al }\right]^2\left[\fr{100}{g^*}\right]^{\fr{1}{3}}S_{sw}(f),
	\label{ome}
	\ee
	where
	\be
	S_{sw}(f)=(f/f_{sw})^{1/3}\left(\frac{7}{4+3(f/f_{peak.sw})^2}\right)^{7/2},
	\ee
	and the peak frequency is well fitted by
	\be
	f_{peak.sw}=1.9\times 10^{-2} mHz \times \frac{1}{v_b}\frac{\beta}{H^*}\frac{T_*}{100}\left(\frac{g_*}{100}\right)^{1/6}.
	\ee
	
GWs from turbulence in the cosmic fluid is given by
	\be
	h^2\Om_{tur}(f) \simeq 3.35\times 10^{-4} \left(\fr{H^*}{\beta}\right) v_b \left(\fr{k_t\al }{1+ \al }\right)^{\fr{3}{2}}\left(\fr{100}{g^*}\right)^{1/3}S_{tur}(f),
	\label{omet}
	\ee
	in which
	\be
	S_{tur}(f)=\frac{(f/f_{peak.tur})^{3}}{\left(1+(f/f_{peak.tur})\right)^{11/3}\left(1+8\pi f/h^*\right)} \label{tur},
	\ee
	and Eq. (\ref{tur}) reads
	\bea
	h_{*} &= &16.5\times 10^{-3} mHz \times \frac{T_*}{100}\left(\frac{g_*}{100}\right)^{1/6}\,,\crn
	f_{peak.tur} &=&2.7\times 10^{-2} mHz \times \frac{1}{v_b}\frac{\beta}{H^*}\frac{T_*}{100}\left(\frac{g_*}{100}\right)^{1/6}.
	\eea
	One can estimate  that \cite{71}
	\bea
	&k_{w}= \al\left[0.73 +0.083\sqrt{\al}+\al\right] \quad\text{when}\quad v_b \sim 1;\\
	&k_t=(0.05-0.1)\times k_{w}.
	\label{k}
	\eea
where $k_w$ is part of the latent heat, converted into kinetic energy of the bubble wall.

It is important to note that the current formulas for the calculation of the gravitational wave spectrum have some differences, possibly slightly from Eqs.~(\ref{ome}) and (\ref{omet}), but they are all the same power. One of the pretty good corrections available today as in Ref.~\cite{grahamgw1}. However, in order for us to only evaluate the power of the gravitational wave spectrum, we only use the formulas in  Eqs.~(\ref{ome}) and (\ref{omet}).

\begin{figure}[h!]
	\centering	
	\begin{subfigure}{.5\textwidth}
		\centering
		\includegraphics[width=1\linewidth]{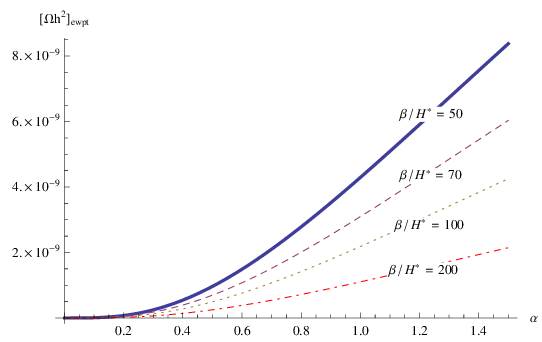}
		\caption{$\Om h^2$ with different $\beta/H^*$ values, $f=f_{peak}$}
		\label{betaalpha}
	\end{subfigure}%
	\begin{subfigure}{.5\textwidth}
		\centering
		\includegraphics[width=1\linewidth]{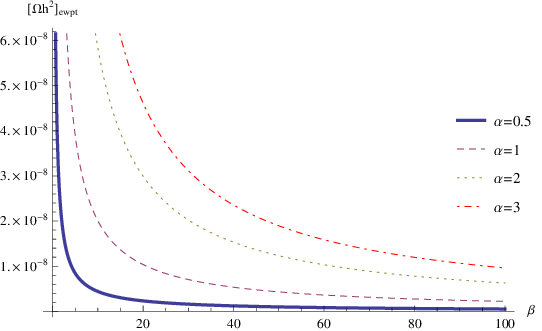}
		\caption{$\Om h^2$ with different $\al $ values, $f=f_{peak}$}
		\label{alphabeta}
	\end{subfigure}
	\caption{The graph of $\Om h^2$ varies with $\al $ and $\beta/H^*$}	
	\label{abf}
\end{figure}

First, we have  to evaluate
 $\al $ and $\beta/H^*$: if $\al $ increases and   $\beta/H^*$ decreases, the density parameter increases. This is shown as Fig.~\ref{abf}.

So, according to Eqs.~(\ref{beta}) and (\ref{all}) if we want the larger density parameter to be, the larger $\al$ is, i.e., the larger change of the potential over temperature should be, and the smaller the change of the sphaleron energy over temperature is at the critical temperature.

By approximating Eq.~(\ref{betah}) and the data in Table \ref{decoupling_table}, we get $\beta/H^*\approx 22.5$. We then calculate $\al$ and the values of $\Om h^2$ with a few $S$, as shown in Table \ref{omegah}.

\begin{table}[ht]
	\begin{center}
		\caption{Gravitational wave corresponding to the strength of phase transition with $\beta/H^{*}=22.5$ and $f=f_{peak}$.}
		\begin{tabular}{c|c||c|c|c|c|c}\hline\hline
			$T_c[\textrm{GeV}]$&$v(T_c)[\textrm{GeV}]$& $\al $&$[\Om h^2]_{sw}$
			&$[\Om h^2]_{tur}$&$\Om h^2]_{ewpt}$& $v_b$ \\ \hline\hline
			
			\multicolumn{7}{c}{$S=1$, $m_{k^{\pm\pm}}=219.496 \,\textrm{GeV}$, $m_{h^{\pm}}=209.293 \,\textrm{GeV} $} \\
			\hline			
			$123.081$&$123.08$&$0.0148$&$6.56499\times10^{-14}$&$2.93648\times10^{-14}$&$3.59298\times10^{-14}$&$0.668$\\ \hline\hline
			\multicolumn{7}{c}{}\\
			\multicolumn{7}{c}{$S=1.2$, $m_{k^{\pm\pm}}=256.169 \,
			\textrm{GeV}$, $m_{h^{\pm}}=218.500\, \textrm{GeV} $}\\
			\hline
			$118.099$&$141.724$&$0.022$&$3.13011\times10^{-14}$&$9.45874\times10^{-14}$&$1.25888\times10^{-13}$&$0.685$\\ \hline\hline
			\multicolumn{7}{c}{}\\
			\multicolumn{7}{c}{$S=1.4$, $m_{k^{\pm\pm}}=294.418\,
			 \textrm{GeV}$, $m_{h^{\pm}}=217.491\, \textrm{GeV}$} \\
			\hline\hline
			$113.606$&$159.049$&$0.0304$&$1.08423\times10^{-13}$&$2.39771\times10^{-13}$&$3.48193\times10^{-13}$&$0.701$\\
			 \hline\hline
			\multicolumn{7}{c}{}\\
			\multicolumn{7}{c}{$S=1.6$,$m_{k^{\pm\pm}}=288.500\, \textrm{GeV}$, $m_{h^{\pm}}=278.355\, \textrm{GeV}$}\\
			\hline
			$108.986$&$174.379$&$0.0403$&$3.26923\times10^{-13}$&$5.47717\times10^{-13}$&$8.7464\times10^{-13}$&$0.717$\\ \hline\hline
			\multicolumn{7}{c}{}\\
			\multicolumn{7}{c}{$S=1.8$, $m_{k^{\pm\pm}}=318.163\,
			 \textrm{GeV}$, $m_{h^{\pm}}=287.461\, \textrm{GeV}$}\\
			\hline
			$104.934$&$188.881$&$0.0505$&$7.81185\times10^{-13}$&$1.05107\times10^{-12}$&$1.83226\times10^{-12}$&$0.730$\\ \hline\hline
		\end{tabular}\label{omegah}
	\end{center}
\end{table}

According to Table \ref{omegah}, as $S$ increases, $v_b$ increases. So the radius of bubbles expressed in Eq.~(\ref{radiusvb}) also increases. We next look at the first and last columns of Table \ref{omegah}, as the temperature decreases, $v_b$ increases again, which tells us that as the temperature decreases the radius of bubbles increases, or the velocity of the walls of bubbles increases, leads to the thickness of the walls can be reduced. This was also commented earlier, when observing the values of the two free parameters $a$ and $b$.

As the temperature decreases, from Tables \ref{decoupling_table} and \ref{omegah} with Eq.~(\ref{radiusvb}), the sphaleron energy increases so the sphaleron rate decreases, consequently the number of bubbles also decreases.

\begin{figure}[h!]
	\centering
	\begin{subfigure}{.5\textwidth}
		\centering
		\includegraphics[width=1\linewidth]{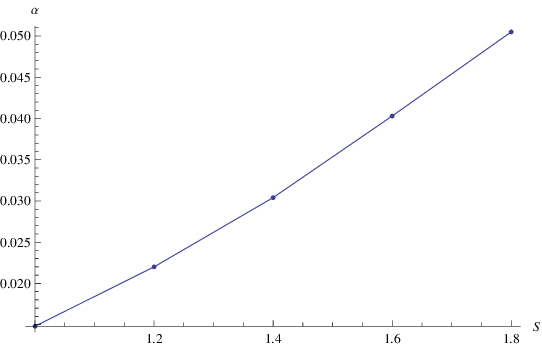}
		\caption{The change of $\al $ with the strength of phase transition.}
		\label{cl-2d}
	\end{subfigure}%
	\begin{subfigure}{.5\textwidth}
		\centering
		\includegraphics[width=1\linewidth]{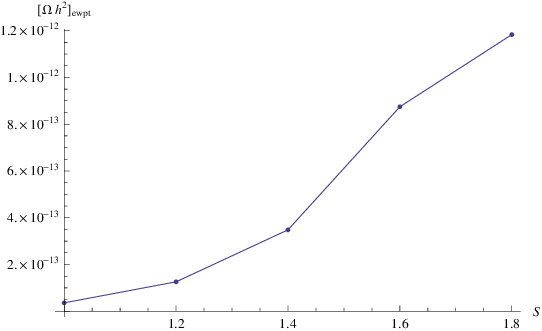}
		\caption{The change of $\Om h^2$ with the strength of phase transition.}
		\label{cl-3d}
	\end{subfigure}
	\caption{Graphs of $\al $ and $\Om h^2$, $f=f_{peak}$.}
	\label{cl-2-3d}
\end{figure}

According to Figs.~\ref{cl-2-3d}, we see that $\Om h^2$ increases when $S$ is greater than $1$ but not yet in the range $10^{-12}-10^{-10}$ and the value of $\al $ increases linearly with $S$. So in the model under consideration, the phase transition strength should be around $3.6$ in the no daisy-loop case (or $S=3.6\times 0.95=3.42$ with daisy loops), $\al$ would have a value in the range of $1$. The last, we conclude that, with the strength of phase transition of about $3.6$, this model shows that the gravitational wave energy density is within the observable region of the LISA. But the maximum strength in the model under consideration is only about $2.4$.

\begin{figure}[h!]
	\centering
	\includegraphics[width=0.7\linewidth]{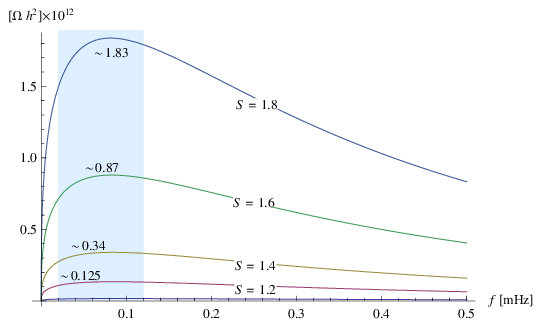}
	\caption{ $\Om h^2$ as a function  of $f$, for different values of $S$. The maximum values are compatible with Table \ref{omegah}. The light blue area represents $f$ from $0.02$ to $0.12$ mHz.}
	\label{gwf1}
\end{figure}

According to Fig.~\ref{gwf1}, the frequency ranges from $2\times 10^{-5}$ Hz to $1.2\times 10^{-4}$ Hz, the value of sensitivity of proposed GW is the largest. The future space-based GW interferometers LISA \cite{Seoane}, DECIGO \cite{kawamura,kudoh} and BBO \cite{harry} has the better sensitivity of proposed GW interferometers as indicated in Table \ref{dulieu}.
	
	\begin{table}[!ht]
		\centering
		\caption{Sensitivity of proposed GW}
		\begin{tabular}{c|c|c|c}\hline\hline
			$f [Hz]$&$\Omega h^2$& Type & Ref.\\ \hline	
			$2\times 10^{-5}-1.2\times 10^{-4}$& $2\times10^{-12}-10^{-10}$ & LISA& Refs.\cite{phu1gw,kudoh,thranel}\\
			$2\times 10^{-5}-1.2\times 10^{-4}$& $5\times 10^{-15}-8\times 10^{-14}$ & DECIGO & Refs.\cite{phu1gw,kudoh,thranel}\\
			$2\times 10^{-5}-1.2\times 10^{-4}$& no data & BBO & Refs.\cite{phu1gw,kudoh,thranel}\\
			\hline\hline
		\end{tabular}\label{dulieu}
	\end{table}	

\begin{figure}[h!]
	\centering
	\includegraphics[width=0.75\linewidth]{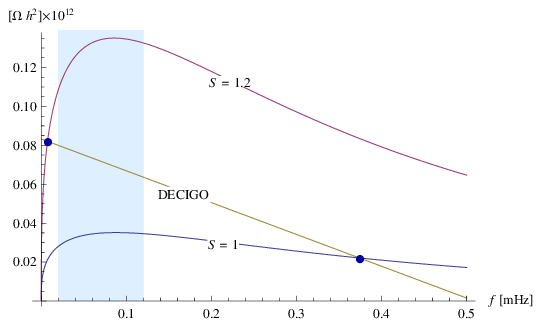}
	\caption{ $\Omega h^2$ as a  function of $f$ for different values of $S$ and the sensitivity of proposed GW of DECIGO \cite{phu1gw,kudoh,thranel}.}
	\label{gwfdecigo}
\end{figure}

From Table \ref{dulieu} and our calculation results, in the future DECIGO can capture the gravitational wave signals generated from EWPT in this model, as see in Fig.~\ref{gwfdecigo}.

\begin{figure}[h!]
	\centering
	\begin{subfigure}{.5\textwidth}
		\centering
		\includegraphics[width=1\linewidth]{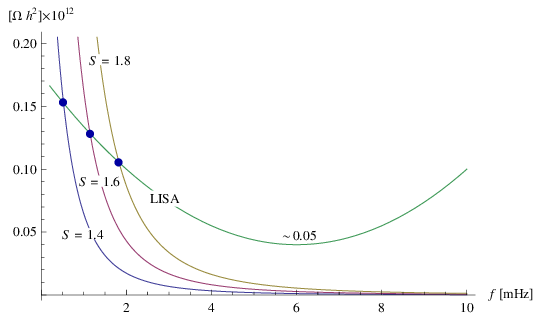}
		\caption{ $\Om h^2$ as a function of $f$ for different values of $S$ and the sensitivity of proposed GW of LISA \cite{phu1gw,kudoh,thranel}.}
		\label{gwfdelisa}
	\end{subfigure}%
	\begin{subfigure}{.5\textwidth}
		\centering
	\includegraphics[width=1\linewidth]{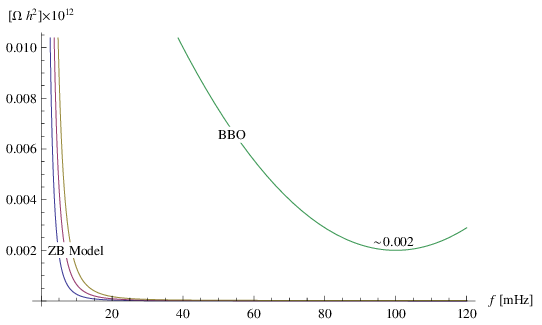}
	\caption{ $\Om h^2$ a function of $f$ for different values of $S$ and the sensitivity of proposed GW of BBO \cite{phu1gw,kudoh,thranel}.}
	\label{gwfbbo}
	\end{subfigure}
	\caption{Graphs of $\Om h^2$ in the ZB model, LISA, BBO.}
	\label{a}
\end{figure}

According to Figs.~\ref{gwfdelisa} and \ref{gwfbbo}, the LISA detectors may detect the GW of EWPT in the frequency range of 2 mHz but the BBO detectors can not.

\section{Conclusion and discussion}\label{sec5}

The sphaleron energy has been  calculated by profile functions with two free parameters. This is a scenario with good results, the numerical solution is not complicated, but the convergence of the energy functional is fast. However, we need to carefully deal with the free constants in the energy functional before solving.

We have noticed that the sphaleron energy needs to satisfy the following, regardless of the way: the first is the de-coupling condition; the second
is the scaling law; the maximum value of sphaleron energy is at zero temperature; and  the last, we can only compute the sphaleron energy in detail at temperature $T\leq T_c$ and can only estimate the rate of
sphaleron at temperatures above $T_c$.

The sphaleron rate tells us about the baryon-number violation, the first condition of baryogenesis. In the ZB model has the nonzero phases of the CKM matrix and a neutrino mixing \cite{74,74b}, they tell us about CP violation.
 The last condition for thermal imbalance is demonstrated by a first order EWPT in this model. Thus, the ZB model allows a baryogenesis scenario that  explains the matter-antimatter asymmetry.

As estimated for $\Om h^2$ to be around $10^{-12}-10^{-10}$ (within the LISA observation limit), $\al $ must be greater than $1$, $S$ must be greater than $3$. However, $S$ cannot be very large (should only be in the vicinity of $2$), otherwise the effective potential expansions are no longer accurate. An error can also come from the slow convergence of perturbation theory in the Quantum Field Theory. This can be greatly improved after the re-summation process as shown in Refs. \cite{grahamgw2, grahamgw2b}.

We also make an extended comment, for looking at the strength of the EWPT  when gravitational waves are included. We have  considered  two parameters $S$ and $\al $. A first-order EWPT is strong when $\al >1$ and $S>1$. So,
in the ZB model, the EWPT is the first-order one but not strong because $\al $ is only about $10^{-2}$. In addition, we have  further investigated the gravitational wave spectrum over time in the phase transition by using the formalism in Ref. \cite{grahamgw1}, but this requires knowing the exact decrease in temperature over time or knowing the details of the expansion of the universe and the growth of bubbles, which must not be approximate.

In this paper, we only calculate the contributions of components $m_{h^\pm}$ and $m_{k^{\pm\pm}}$. However the full masses of $h^{\pm}$ and $k^{\pm\pm}$ are Eq.~(\ref{masshk}), so that their masses still depend on $u_1$ and $u_2$.

To comply with some two-loop effects, for neutrino mass generation, the masses of these new bosons are not allowed to exceed 2 TeV \cite{test}. In the Zee-Babu model, we can have the decay $k^{\pm\pm}\longrightarrow h^{\pm}h^{\pm}$ and $k^{\pm\pm}\longrightarrow \tau^{\pm}\tau^{\pm}$ is very small. Therefore, $M_{k{\pm\pm}} > 220$ GeV and $2M_{h^\pm}<M_{k^{\pm\pm}}$ \cite{test}. With the results in this paper, to have the first order phase transition and the sphaleron energies are in the range of 5-10 TeV, their massed are about less than 300 GeV. However, this does not violate the above conditions because we have not taken into account $u_1$ and $u_2$.
$u_1$ and $u_2$ are the parameters that go with the new charged bosons. These coupling constants are not the coupling constants between two new bosons with Higgs. Therefore, the EWPT calculation will not have the contribution of these two parameters. $m_{h} (m_k)$ participates in EWPT, not $M_h (M_k)$. So when calculating the 1-loop effective potential, these two parameters do not affect the sphaleron and GW.

The recent new effects at the LHC have shown signals of di-bosons that have masses around 2 TeV \cite{atlas, atlasb}. The new bosons in the Zee-Babu model participate in the electroweak process with only a part of their masses. The other mass part comes from $u_1, u_2$. As calculated for the first order EWPT, the part involved in this phase transition is just under $300$ GeV. But the mass of these new bosons can be up to TeV due to the contributions of $u_1, u_2$. Therefore, although the energy scale is about TeV, it is too high for the EWPT scale. But if the signals of these new particles are confirmed to be particles in the ZB model, then as the TeV scale goes down the EWPT scale, the part of these new particles  involves in the electroweak process.

Another very interesting suggestion comes from the analysis of the decays of doubly charged Higgs boson in the minimal left-right symmetry model when it comes to FCNC effects and vacuum stability of the scalar potential \cite{bambhaniya}. The decays $p p \rightarrow H_{1/2}^{\pm \pm} H_{1/2}^{\mp \mp} j j \rightarrow \ell^{\pm} \ell^{\pm} \ell^{\mp} \ell^{\mp} jj$ with $\sqrt{s}=14 \, \textrm{TeV}$, LHC2 will not detect signals $H_{1/2}^{\pm\pm}H_{1/2}^{\mp\mp}jj$, even with light doubly charged Higgs boson ($\sim 500\, \mathrm{GeV}$). But the Future Circular Colliders with $\sqrt{s}=33$ and (or) 100 TeV, doubly charged Higgs bosons with a mass of about 1 TeV can be easily detected. The doubly charged Higgs bosons in the Zee-Babu model are used to generate the small mass of neutrinoes at the 2 loop level. Therefore, their signals in the above decays with $\sqrt{s}\ge 33 \, \textrm{TeV}$ can also be detected. This is an open suggestion that needs careful analysis.

An interesting effect in this model is the small mass generation of neutrino at the two-loop level. We think that this two-loop effect will have a certain influence on the EWPT. Therefore, we will calculate the effective potential at the two-loop to examine the baryogenesis problem fully in the ZB model.

\section*{Acknowledgements}

In memory of the day VQP's venerable teacher, Dr. Vo Thanh Van, passed away on 5 July 2016, this series of our papers highlights the continuation of what he has done through many years in the Department of Theoretical Physics. We would like
to express our sincere gratitude to the referee for useful remarks leading to  this version.


\begin{thebibliography}{99}
	
\bibitem{72} M.Fukugita and Yanagida, Phys.\ Lett.\ B {\bf 174}, 45 (1986)

\bibitem{sakharov} A. D. Sakharov, JETP Lett. \textbf{5}, 24 (1967)

\bibitem{mkn} M. I. Dine, R. G. Leigh, P. Huet, A. Linde, and D. Linde, Phys. Rev. D \textbf{46}, 550 (1992)

\bibitem{SME}K. Kajantie, M. Laine, K. Rummukainen, and M. Shaposhnikov, Phys. Rev. Lett. \textbf{77}, 2887 (1996)

\bibitem{SMEb} F. Csikor, Z. Fodor, and J. Heitger, Phys. Rev. Lett. \textbf{82}, 21 (1999)

\bibitem{SMEc} J. Grant, M. Hindmarsh, Phys. Rev. D \textbf{64}, 016002 (2001)

\bibitem{SMEd} M. D'Onofrio, K. Rummukainen, A.  Tranberg, JHEP \textbf{08}, 123 (2012)

\bibitem{michela} M. D'Onofrio, K. Rummukainen, A. Tranberg, Phys. Rev. Lett. \textbf{113}, 141602 (2014)

\bibitem{2} Vo Quoc Phong, Vo Thanh Van, and Hoang Ngoc Long, Phys. Rev. D \textbf{88}, 096009 (2013), arXiv:1309.0355[hep-ph]

\bibitem{2b} A. Menon, D. E. Morrissey, and C. E. M. Wagner, Phys. Rev. D \textbf{70}, 035005 (2004)

\bibitem{2c} S. W. Ham, S. K. Oh, C. M. Kim, E. J. Yoo, and D. Son, Phys. Rev. D \textbf{70}, 075001 (2004)

\bibitem{BSM}M. Bastero-Gil, C. Hugonie, S. F. King, D. P. Roy, and S. Vempati, Phys. Lett. B \textbf{489}, 359 (2000)

\bibitem{BSMb} A. Menon, D. E. Morrissey, and C. E. M. Wagner, Phys. Rev. D \textbf{70}, 035005 (2004)

\bibitem{BSMc} S. W. Ham, S. K. Oh, C. M. Kim, E. J. Yoo, and D. Son, Phys. Rev. D \textbf{70}, 075001 (2004)

\bibitem{majorana} J.  M. Cline, G.  Laporte, H.  Yamashita, S. Kraml, JHEP \textbf{0907}, 040 (2009)

\bibitem{majoranab} Aleksandr Azatov, Miguel Vanvlasselaer, JHEP \textbf{09}, 085 (2020)

\bibitem{thdm} S. Kanemura, Y. Okada, E. Senaha, Phys. Lett. B \textbf{606}, 361-366 (2005)

\bibitem{thdmb} G. C. Dorsch, S. J. Huber, J. M. No, JHEP \textbf{10}, 029 (2013)

\bibitem{ESMCO} S. W. Ham, S-A Shim, and S. K. Oh, Phys. Rev. D \textbf{81}, 055015 (2010)

\bibitem{elptdm} D. Borah and J. M. Cline, Phys. Rev. D \textbf{86}, 055001 (2013)

\bibitem{elptdma} A. Ahriche and S. Nasri, Phys. Rev. D \textbf{85}, 093007 (2012)

\bibitem{elptdmb} S.  Das, P.  J. Fox, A. Kumar, and N. Weiner, JHEP \textbf{1011}, 108 (2010)

\bibitem{elptdmc} D. Chung and A. J. Long, Phys. Rev.  D \textbf{84}, 103513 (2011)

\bibitem{elptdmd} M.  Carena, N. R. Shaha, and C. E. M. Wagner, Phys. Rev. D \textbf{85}, 036003 (2012)

\bibitem{phonglongvanb} J. Sá Borges, R. O.Ramos, Eur. Phys. J. C \textbf{76}, 344 (2016)

\bibitem{phonglongvan2} V. Q.  Phong,  H.  N. Long, V.  T.  Van, L. H.  Minh, Eur. Phys. J. C \textbf{75}, 342 (2015), arXiv:1408.5657[hep-ph]

\bibitem{SMS} J. R. Espinosa, T. Konstandin and F. Riva, Nucl. Phys. B \textbf{854}, 592 (2012)

\bibitem{dssm}S. Kanemura, E.  Senaha, T.  Shindou and T.  Yamada, JHEP \textbf{1305}, 066 (2013)

\bibitem{munusm}D. J. H. Chung and A.  J.  Long, Phys. Rev. D \textbf{81}, 123531 (2010)

\bibitem{lr} G. Barenboim and N. Rius, Phys. Rev. D \textbf{58}, 065010 (1998)

\bibitem{singlet} S. Profumo, M. J. Ramsey-Musolf, G. Shaughnessy, JHEP \textbf{0708}, 010 (2007)

\bibitem{singletb} S. Profumo, M. J. Ramsey-Musolf, C. L. Wainwright, P. Winslow, Phys. Rev. D \textbf{91}, 035018 (2015)

\bibitem{singletc} D. Curtin, P. Meade, C-T. Yu, JHEP \textbf{11}, 127 (2014)

\bibitem{singletd} M. Jiang, L. Bian, W. Huang, J. Shu, Phys. Rev. D \textbf{93}, 065032 (2016)

\bibitem{mssm1} M. Carena, G. Nardini, M. Quiros, C. E.M. Wagner, Nucl. Phys. B \textbf{812}, 243-263 (2009)

\bibitem{mssm1b} A. Katz, M. Perelstein, M. J. Ramsey-Musolf, P. Winslow, Phys. Rev. D \textbf{92}, 095019 (2015)

\bibitem{mssm1c} J. Kozaczuk, S. Profumo, L. S. Haskins, C. L. Wainwright, JHEP \textbf{1501}, 144  (2015)

\bibitem{twostep} H. H. Patel, M. J. Ramsey-Musolf, Phys.Rev. D. \textbf{88}, 035013 (2012)

\bibitem{twostepb} N. Blinov, J. Kozaczuk, D. E. Morrissey, C. Tamarit, Phys. Rev. D \textbf{92}, 035012 (2015)

\bibitem{twostepc} S. Inoue, G. Ovanesyan, M. J. Ramsey-Musolf, Phys. Rev. D \textbf{93}, 015013 (2016)

\bibitem{1101.4665} H. H. Patel, M. J. Ramsey-Musolf, JHEP \textbf{1107}, 029 (2011)

\bibitem{1101.4665b} G. W. Anderson and L. J. Hall, Phys. Rev. D \textbf{45}, 2685 (1992)

\bibitem{1101.4665c} H. H.Patel and M.J.Ramsey-Musolf, M. Garny and T.Konstandin, JHEP \textbf{1207}, 189 (2012)

\bibitem{jjgb} J. De Vries, M. Postma, J. van de Vis, JHEP \textbf{1904}, 024 (2019)

\bibitem{jjgc} J. de Vries, M. Postma, J. van de Vis , G. White, JHEP \textbf{1801}, 089 (2018)

\bibitem{jjgd} C. Balazs, G. White, J. Yue, JHEP \textbf{1703}, 030 (2017)

\bibitem{Ahriche1} A.~Ahriche, Phys.\ Rev.\ D {\bf 75}, 083522 (2007)

\bibitem{Ahriche2} A.~Ahriche, Eur.\ Phys.\ J.\ C {\bf 66}, 333 (2010)

\bibitem{Ahriche2b} T.~A.~Chowdhury and S.~Nasri, JHEP {\bf 1411}, 096 (2014)

\bibitem{Ahriche3} A.~Ahriche and S.~Nasri, JCAP {\bf 1307}, 035 (2013)

\bibitem{Ahriche3b} A.~Ahriche, G.~Faisel, S.~Y.~Ho, S.~Nasri and J.~Tandean, Phys.\ Rev.\ D {\bf 92}, 035020 (2015)

\bibitem{Ahriche3c} A.~Ahriche, K.~L.~McDonald and S.~Nasri, Phys.\ Rev.\ D {\bf 92}, 095020 (2015)

\bibitem{Ahriche3d} A.~Ahriche, S.~M.~Boucenna and S.~Nasri, Phys.\ Rev.\ D {\bf 93}, 075036 (2016)

\bibitem{Ahriche3e} A.~Ahriche, K.~Hashino, S.~Kanemura and S.~Nasri, Phys.\ Lett.\ B {\bf 789}, 119 (2019)

\bibitem{Fuyuto} K.~Fuyuto and E.~Senaha, Phys.\ Rev.\ D {\bf 90}, 015015 (2014)

\bibitem{Fuyutob} K.~Fuyuto and E.~Senaha, Phys.\ Lett.\ B {\bf 747}, 152 (2015)

\bibitem{Fuyutoc} K.~Funakubo and E.~Senaha, Phys.\ Rev.\ D {\bf 79}, 115024 (2009)

\bibitem{span} M.~Spannowsky and C.~Tamarit, Phys.\ Rev.\ D {\bf 95}, 015006 (2017)

\bibitem{chr} C. Grojean, G. Servant, J. D. Wells, Phys. Rev. D \textbf{71} 036001 (2005)

\bibitem{cde} C. Delaunay, C. Grojean, J. D. Wells, JHEP \textbf{0804}, 029 (2008)

\bibitem{kusenko} A. Kusenko, L. Pearce, and L. Yang, Phys. Rev. Lett.\textbf{ 114}, 061302 (2015)

\bibitem{epjc} V. Q.  Phong,  H.  N. Long, V.  T.  Van, L. H.  Minh, Eur. Phys. J. C \textbf{75}, 342 (2015), arXiv:1409.0750[hep-ph]

\bibitem{chiang3} C. W. Chiang, T. Yamada, Phys. Let. B \textbf{735}, 295 (2014)

\bibitem{zb} V. Q. Phong, N. C. Thao, H. N. Long, Phys. Rev. D \textbf{97}, 115008 (2018)

\bibitem{Arefe} J. R. Espinosa, T. Konstandin, J. M. No and M. Quiros, Phys. Rev. D \textbf{78}, 123528 (2008)

\bibitem{r23} D. Comelli, J.R. Espinosa, Phys.Rev. D \textbf{55}, 6253 (1997)

\bibitem{spha-huble} M. Joyce, Phys. Rev. D \textbf{55}, 1875 (1997)

\bibitem{decoupling} K. Funakubo and E. Senaha, Phys. Rev. D \textbf{79}, 115024 (2009)

\bibitem{decouplingb} M. Dvornikov and V. B. Semikoz, Phys. Rev. D \textbf{87}, 025023 (2013)

\bibitem{decouplingc} T.  M. Gould and I. Z. Rothstein, Phys. Rev. D \textbf{48}, 5917 (1993)

\bibitem{decouplingd} K. Fuyuto and E. Senaha, Phys. Rev. D \textbf{90}, 015015 (2014)

\bibitem{47} S. Braibant, Y. Brihaye and J. Kunz, Int. J.  Mod. Phys. A \textbf{08}, No. 31, 5563 (1993)

\bibitem{hagi} K. Hagiwara, S. Ishihara, R. Szalapski and D. Zeppenfeld, Phys. Rev. D \textbf{48}, 2182 (1993)

\bibitem{10} F. R. Klinkhamer and N. S. Manton, Phys. Rev. D \textbf{30}, 2212 (1984)

\bibitem{phu2spha} M. Spannowsky and C. Tamarit, Phys. Rev. D \textbf{95}, 015006 (2017)

\bibitem{phu1spha} S. Das, P. J. Fox, A. Kumar, and N. Weiner, JHEP \textbf{1011}, 108 (2010)

\bibitem{phu3spha} F. R. Klinkhamer and P. Nagel, Phys. Rev. D \textbf{96}, 016006 (2017)

\bibitem{phu4spha} A. Friedlander, I. Banta, J. M. Cline, and D. Tucker-Smith, Phys. Rev. D \textbf{103}, 055020 (2021)

\bibitem{13} S. W. Ham and S. K. Oh, Phys. Rev. D\textbf{ 70}, 093007 (2004)

\bibitem{11} X. Gan, A. J. Long, and L-T. Wang, Phys. Rev. D \textbf{96}, 115018 (2017)

\bibitem{gauge0} Y. Brihaye and J. Kunz, Phys. Rev. D \textbf{48}, 3884 (1993)

\bibitem{cosmotran1b} C. L. Wainwright, Comput. Phys. Commun. \textbf{183}, 2006 (2012)

\bibitem{plk} Vo Quoc Phong, Phan Hong Khiem, Ngo Phuc Duc Loc, Hoang Ngoc Long, Phys. Rev. D \textbf{101}, 116010 (2020),  arXiv:2003.09625[hep-ph]

\bibitem{1gw} Kip S. Thorne, \textit{Gravitational Waves}, arXiv:gr-qc/9506086v1 (1995)

\bibitem{2gw} L. Leitao, A. Megevand, A. D. Sanchez, JCAP \textbf{10}, 024 (2012)

\bibitem{2gwa} M. S Tuner and S. Wilzek, Phy. Rev. \textbf{65}, 3080 (1990)

\bibitem{2gwb} Christophe Grojean, G. Servant, Phys. Rev. D \textbf{75}, 043507 (2007)

\bibitem{phu1gw} S.V. Demidov, D. S. Gorbunov, D. V. Kirpichnikov, Physics Letters B \textbf{779}, 191 (2018).

\bibitem{phu2gw} S. J. Huber, T. Konstandin, JCAP \textbf{0809}, 022 (2008).

\bibitem{phu2gwb} Ryusuke Jinno, Masahiro Takimoto, Phys. Rev. D \textbf{95}, 024009 (2017)

\bibitem{phu3gw} D. J. H. Chung, and P. Zhou, Phys. Rev. D \textbf{82}, 024027 (2010)

\bibitem{phu4gw} Zh. Kang, JHEP \textbf{02}, 115 (2018)

\bibitem{phu5gw} R. Zhou, L. Bian and H-K Guo, Phys. Rev. D \textbf{101}, 091903 (2020)

\bibitem{phu6gwsimulation} B. Laurent and J. M. Cline, Phys. Rev. D \textbf{102}, 063516 (2020)

\bibitem{phu7gw} J. M. Cline, A. Friedlander, D-M He, K. Kainulainen, B. Laurent, D. Tucker-Smith, Phys. Rev. D \textbf{103}, 123529 (2021)

\bibitem{phu8gw} O. Gould, J. Kozaczuk, L. Niemi, M. J. Ramsey-Musolf, T. V. I. Tenkanen, D. J. Weir, Phys. Rev. D \textbf{100}, 115024 (2019)

\bibitem{3gw} M. F. Axen, S. Banagiri, A. Matas, C. Caprini, V. Mandic, Phys. Rev. D \textbf{98}, 103508 (2018)

\bibitem{4gw} M. Kamionkowski, A. KosowsKy and M. Turner,Phys. Rev. D\textbf{ 49}, 2837 (1994)

\bibitem{8gw} A. D. Dolgov, D. Grasso, A. Nicolis, Phys. Rev. D \textbf{66}, 103505 (2002)

\bibitem{9gw} K. Schmitz, JHEP \textbf{01}, 097 (2021)

\bibitem{9gwb} R. Apredaa, M. Maggioreb, A. Nicolisc and A. Riotto, Nucl. Phys. B \textbf{631}, 342-368 (2002)

\bibitem{7gw} C. Grojean and G. Servant, Phys. Rev. D \textbf{75}, 043507 (2007)

\bibitem{7gwb} M. Maggiore, Phys. Rept. \textbf{331}, 283-367 (2000)

\bibitem{zeebabu} A. Zee, Nucl. Phys. B \textbf{264}, 99 (1986)

\bibitem{zeebabub} K. S. Babu, Phys. Lett. B \textbf{203}, 132 (1988)

\bibitem{bzee} H.  Okada, T. Toma, K.  Yagyu, Phys. Rev. D \textbf{90}, 095005 (2014)

\bibitem{bzeea} S.  Baek, JHEP \textbf{08}, 023 (2015)

\bibitem{bzeeb} H.  Okada, Y.  Orikasa, Phys. Lett. B \textbf{760}, 558 (2016)

\bibitem{carrington} M. E. Carrington, Phys. Rev. D \textbf{45}, 2933 (1992)

\bibitem{curtin} David Curtin, Patrick Meade, Harikrishnan Ramani, Eur. Phys. J. C \textbf{78}, 787 (2018)

\bibitem{katz} A. Katz and M. Perelstein, JHEP \textbf{07} 108 (2014)

\bibitem{espinosa} J.R. Espinosa, M. Quiros and F. Zwirner, Phys. Lett. B \textbf{291}, 115 (1992).

\bibitem{garcia}  Juan Herrero-Garcia, Miguel Nebot, Nuria Rius, Nucl. Phys. B \textbf{885}, 542-570 (2014)

\bibitem{5percent} G.  W. Anderson and L.  J. Hall, Phys. Rev. D \textbf{45}, 2685 (1992)

\bibitem{71} Chiara Caprini, Mark Hindmarsh, Stephan Huber, Thomas Konstandin, Jonathan Kozaczuk, Germano Nardini, Jose Miguel No, Antoine Petiteau, Pedro Schwaller, Geraldine Servant, David J. Weir, JCAP\textbf{04}, 001 (2016)

\bibitem{73} A. Kosowsky, M. S. Turner, and R. Watkins, Phys. Rev. D \textbf{45}, 4514 (1992)

\bibitem{73a} A. Kosowsky, M. S. Turner, and R. Watkins, Phys. Rev. Lett. \textbf{69}, 2026 (1992)

\bibitem{73b} A. Kosowsky and M. S. Turner, Phys. Rev. D \textbf{47}, 4372 (1993)

\bibitem{73c} S. J. Huber and T. Konstandin, JCAP \textbf{08099}, 022 (2008)

\bibitem{73d} R. Jinno and M. Takimoto, Phys. Rev. D \textbf{95}, 024009 (2017)

\bibitem{73e} R. Jinno and M. Takimoto, JCAP \textbf{1901}, 060 (2019)

\bibitem{grahamgw1} Huai-Ke Guo, K. Sinha, D. Vagie, G. White, 	JCAP \textbf{01}, 001 (2021)

\bibitem{Seoane} P.A. Seoane, et al., eLISA Collaboration, arXiv:1305.5720 [astro-ph.CO]

\bibitem{kawamura} N. Seto, S. Kawamura, T. Nakamura, Phys. Rev. Lett. \textbf{87}, 221103 (2001)

\bibitem{kudoh} Hideaki Kudoh, Atsushi Taruya, Takashi Hiramatsu, and Yoshiaki Himemoto, Phys. Rev. D \textbf{73}, 064006 (2006)

\bibitem{harry} G.M. Harry, P. Fritschel, D.A. Shaddock, W. Folkner, E.S. Phinney, Class. Quantum Gravity \textbf{23}, 4887 (2006); Erratum:
Class. Quantum Gravity \textbf{23}, 7361 (2006).

\bibitem{thranel} Eric Thrane1 and Joseph D. Romano, Phys. Rev. D \textbf{88}, 124032 (2013)

\bibitem{74} T. Ohlsson, T. Schwetz, He Zhang, Phys. Lett. B \textbf{681}, 269 (2009)

\bibitem{74b} Hoang Ngoc Long and Vo Van Vien, 	Int. J. Mod. Phys. A \textbf{29}, 1450072 (2014)

\bibitem{grahamgw2} Huai-Ke Guo, K. Sinha, D. Vagie, G. White, JHEP\textbf{06}, 164 (2021)

\bibitem{grahamgw2b} D. Croon, O. Gould, P. Schicho, T. V. I. Tenkanen, G. White, JHEP \textbf{04}, 055 (2021)

\bibitem{test}J.  Herrero-Garcia, M.  Nebot, N.  Rius and A.  Santamaria, Nucl. Phys. B \textbf{885}, 542-570 (2014)

\bibitem{atlas}G. Aad et al, [ATLAS Collaboration], JHEP \textbf{12}, 55 (2015)

\bibitem{atlasb} V. Khachatryan et al [CMS Collaboration], JHEP \textbf{1408}, 173 (2014)

\bibitem{bambhaniya} G. Bambhaniya, J. Chakrabortty, J. Gluza, T. Jelinski, R. Szafron, Phys. Rev. D \textbf{92}, 015016 (2015)
\end{thebibliography}
\end{document}